\documentclass[twocolumn]{aastex631}

\newcommand{\sersic}{S\'{e}rsic}

\newcommand{\SII}{[\rm{S}\,\textsc{ii}]}
\newcommand{\OIII}{[\rm{O}\,\textsc{iii}]}
\newcommand{\NII}{[\rm{N}\,\textsc{ii}]}
\newcommand{\GALFIT}{\texttt{GALFIT}}
\newcommand{\Phot}{\texttt{Photutils}}

\shorttitle{Disk Galaxies with Double-lobed Radio-loud AGNs}
\shortauthors{Wu, Ho \& Zhuang}
\graphicspath{{./}{figures/}}
\submitjournal{ApJ}
\begin{document}

\title{An Elusive Population of Massive Disk Galaxies Hosting Double-lobed Radio-loud AGNs}

\correspondingauthor{Zihao Wu}
\email{zhwu@pku.edu.cn}

\author[0000-0002-8876-5248]{Zihao Wu}
\affiliation{Department of Astronomy, School of Physics, Peking University, Beijing 100871, People's Republic of China}

\author[0000-0001-6947-5846]{Luis C. Ho}
\affiliation{Kavli Institute for Astronomy and Astrophysics, Peking University, Beijing 100871, People's Republic of China}
\affiliation{Department of Astronomy, School of Physics, Peking University, Beijing 100871, People's Republic of China}

\author[0000-0001-5105-2837]{Ming-Yang Zhuang}
\affiliation{Department of Astronomy, School of Physics, Peking University, Beijing 100871, People's Republic of China}
\affiliation{Kavli Institute for Astronomy and Astrophysics, Peking University, Beijing 100871, People's Republic of China}

\begin{abstract}
It is commonly accepted that radio-loud active galactic nuclei are hosted exclusively by giant elliptical galaxies. We analyze high-resolution optical Hubble Space Telescope images of a sample of radio galaxies with extended double-lobed structures associated with disk-like optical counterparts. After systematically evaluating the probability of chance alignment between the radio lobes and the optical counterparts, we obtain a sample of 18 objects likely to have genuine associations. The host galaxies have unambiguous late-type morphologies, including spiral arms, large-scale dust lanes among the edge-on systems, and exceptionally weak bulges, as judged by the low global concentrations, small global \sersic\ indices, and low bulge-to-total light ratios (median $B/T = 0.13$). With a median \sersic\ index of 1.4 and low effective surface brightnesses, the bulges are consistent with being pseudo bulges. The majority of the hosts have unusually large stellar masses (median $M_* = 1.3\times 10^{11}\, M_\odot$) and red optical colors (median $g-r = 0.69$\,mag), consistent with massive, quiescent galaxies on the red sequence. We suggest that black hole mass (stellar mass) plays a fundamental role in launching large-scale radio jets, and that the rarity of extended radio lobes in late-type galaxies is the consequence of the steep stellar mass function at the high-mass end. The disk radio galaxies have mostly Fanaroff-Riley type~II morphologies yet lower radio power than sources of a similar type traditionally hosted by ellipticals. The radio jets show no preferential alignment with the minor axis of the galactic bulge or disk, apart from a possible mild tendency for alignment among the most disk-dominated systems.
\end{abstract}

\section{Introduction} \label{sec:intro}

Massive black holes (BHs) lurk at the center of all massive galaxies \citep{Kormendy_2013_ARAA_51_511} and even in a significant fraction of less massive systems \citep{Greene_2020_ARAA_58_257}. Mass accretion onto the BH generates nuclear activity that gives rise to a plethora of energetic phenomena associated with active galactic nuclei (AGNs). One of the most enduring unsolved problems is the origin of radio jets, a distinctive feature of some but not all AGNs \citep{Urry_1995_PASP_107_803}. AGNs are conventionally designated as either radio-loud or radio-quiet, even if there is strong motivation to abandon these labels in favor of more physical alternatives (for a recent perspective, see \citealt{Padovani2017}). For historical reasons, two operational definitions of radio-loudness have been commonly invoked. On the one hand, radio-loudness can be demarcated on the basis of an absolute radio power, such as $P_{\rm 6\,cm} > 10^{25}\,\rm W\,Hz^{-1}\,sr^{-1}$ \citep{Miller_1990_MNRAS}. This has the advantage of simplicity and independence from observations at other wavelengths. On the other hand, it has become popular to define radio-loudness not by an absolute but instead by the relative strength of the radio emission. For instance, \cite{Kellermann1989} proposed the widely embraced convention of $R \equiv L_{\nu}({\rm 6\,cm})/L_{\nu}(B)$, with the boundary between radio-loud versus radio-quiet set at $R = 10$. By this criterion, $\sim 10\%-20\%$ of optically selected quasars are radio-loud (e.g., \citealt{Visnovsky1992, Hooper1995}). Both of these traditional criteria encounter difficulties once we confront AGNs of lower luminosity drawn from BHs of lower mass or lower accretion rate. For the vast majority of the galaxy population and for most of their lifecycle, the radio-loudness of an AGN cannot be measured accurately without taking into account host galaxy contamination (e.g., \citealt{Ho_Peng_2001}). Notwithstanding these complications, a distinctive empirical trend regarding the nature of radio jets has remained largely intact: the most powerful radio-loud AGNs, especially those that sport large-scale, double-lobed hotspots that extend beyond the confines of the host galaxy, reside nearly universally in elliptical galaxies, while the host galaxies of the radio-quiet counterparts span a wide range of optical morphologies \citep[e.g.,][]{McLure_1999_MNRAS_308, Kim2017}. The dichotomy in the host morphology in terms of radio-loudness has long been recognized in radio galaxies \citep{Matthews_1964_ApJ_140_35, Zirbel_1996_ApJ, McLure_2004_MNRAS_351_347} and confirmed by high-resolution observations of the hosts of radio-loud quasars \citep{McLure_1999_MNRAS_308, Hamilton_2002_ApJ_576_61, Dunlop_2003_MNRAS_340_1095} and BL~Lac objects \citep{Urry_2000_ApJ_532_816}. Throughout this paper, we restrict our attention to the class of radio-loud AGNs typified by extended, double-lobed radio jets.

What is the physical basis for this observational trend? The leading model suggests that powerful, relativistic jets in radio-loud AGNs form by tapping into the rotational energy of a rapidly spinning BH \citep{Blandford_1977_MNRAS_179_433}. In an attempt to link radio-loudness with galaxy morphology, \cite{Wilson_1995_ApJ_438_62} proposed that radio-loud AGNs are uniquely associated with elliptical galaxies because their central BHs have been spun up by galaxy-galaxy major mergers. Spiral galaxies, on the other hand, cannot spin up BHs because they evolve largely through internal secular processes. Wilson \& Colbert's merger-driven mechanism of generating BH spin explains why spiral galaxies cannot host radio-loud AGNs. \cite{Baum_1995_ApJ} further suggested that the spin paradigm may further be able to account for the difference in radio power between \citet[][FR]{Fanaroff_1974_MNRAS_167_31P} type~I and type~II radio galaxies: FR~IIs are more powerful than FR~Is because they have more rapidly spinning BHs.

Recent advances in BH spin measurements have severely challenged this spin paradigm. Although the number of supermassive BHs with reliable spin measurements is still quite limited, the extant evidence suggests that many supermassive BHs, at least in the nearby Universe, are rapidly spinning objects \citep{Reynolds_2019_Nat_3_41, Reynolds_2021_ARAA_59}, with a significant fraction hosted by spiral galaxies \citep[e.g.,][]{Marinucci_2014_ApJ_787_83, Walton_2014_ApJ_788_76, Vasudevan_2016_MNRAS_458_2012, Buisson_2018_MNRAS_480_3689, Jiang_2019_MNRAS_483_2958}. Notably, recent observations from the Event Horizon Telescope \citep{Akiyama_2022_ApJ} reveal that the central BH of our own Galaxy, Sgr~A*, itself has a high spin. In light of these developments, it is no longer tenable to invoke the spin paradigm to account for the morphology dichotomy between radio-loud and radio-quiet AGNs.

Perhaps BH mass, not spin, determines jet power. After all, the fraction of galaxies that hosts radio-loud AGNs depends strongly on stellar mass \citep{Scarpa_2001_ApJ_556_749,Best_2005_MNRAS_362_25,Mauch_2007_MNRAS_375_931}. Now, to the extent that BH mass is closely linked with stellar mass \citep{Magorrian1998, Kormendy_2013_ARAA_51_511}, it is tempting to conclude that more massive BHs are more prone to producing radio-loud AGNs. Moreover, the very power of the radio jet increases with BH mass \citep[e.g.,][]{Franceschini_1998_MNRAS_297_817, Lacy_2001_ApJL_551_L17, McLure_2004_MNRAS_353_L45}. As late-type galaxies generally host less massive BHs \citep{Greene_2020_ARAA_58_257}, likely an indirect manifestation of their typically lower stellar masses \citep{Moffett2016}, it would be natural that disk-dominated galaxies can only sustain less powerful, and hence less extended \citep{Ledlow2002}, radio jets. These expectations are merely indicative, at best, for they are based on statistical correlations with substantial scatter. Given the significant dispersion in the relation between BH mass and radio luminosity \citep[e.g.,][]{Ho_2002_ApJ_564_120, Woo_2002_ApJ_579_530} and the similarly loose correlation between BH mass and galaxy stellar mass for late-type galaxies \citep{Greene_2020_ARAA_58_257}, we should remain open to the possibility that some late-type galaxies might host radio-loud AGNs.

What concrete evidence exists, then, that disk galaxies host radio-loud AGNs? Much attention has been cast on the so-called radio-loud narrow-line Seyfert 1 galaxies (NLS1s). NLS1s designate AGNs with relatively low BH masses and high accretion rates (e.g., \citealt{Boroson2002}), which normally tend to be radio-quiet \citep{Ulvestad1995, Greene2006}. Consistent with their low BH masses, the hosts of NLS1s are generally disk galaxies (e.g., \citealt{Orban_de_Xivry_2011, Kim2017}). A minority of NLS1s, however, exhibit blazar-like characteristics at radio and even $\gamma$-ray energies (e.g., \citealt{Abdo2009, Foschini_2015_AAP_575_A13}), strongly suggesting that at least in this sub-population late-type galaxies can launch relativistic jets. If true, this would represent a fundamental paradigm shift. However, despite numerous efforts to characterize the structure and morphology of the host galaxies of radio-loud NLS1s, disentangling the host from the glare of the bright nucleus remains challenging using ground-based images, even under conditions of exceptionally good seeing and pushing the observing bandpass to the near-infrared. While some examples of disk hosts have emerged (e.g., \citealt{Kotilainen2016, Vietri_2022_arXiv2204.00020}), others are still open to interpretation (e.g., \citealt{Leon-Tavares2014, Jarvela2018, Olguin-Iglesias2020}), raising doubts as to whether they are bona fide late-type disk galaxies \citep{Tadhunter_2016_AAPR_24_10}. The situation is considerably simpler in the case of radio galaxies, whose orientation in the plane of the sky shields the bright nucleus from the viewer. To date, only a handful of cases of extended radio lobes have been reported to be hosted by disk galaxies \citep{Ledlow_1998_ApJ_495_227, Hota_2011_MNRAS_417_L36, Bagchi_2014_ApJ_788_174, Mao_2015_MNRAS_446_4176, Singh_2015_MNRAS_454_1556, Mulcahy_2016_AAP_595_L8, Vietri_2022_arXiv2204.00020}. However, most discoveries lack detailed chance alignment analysis to confirm the radio associations, and the number of objects is too marginal for meaningful statistical analysis. 

In this study, we present a sample of radio-loud AGNs with classical double-lobed radio structures that we can associate with high confidence with host galaxies having clearly late-type, disk-dominated optical morphology. These objects originate from a parent sample of objects initially selected by the Gems of Galaxy Zoo \citep[Zoo Gems;][]{Keel_2022_AJ_163_150} project, which obtained high-resolution Hubble Space Telescope (HST) optical images of a collection of sources previously identified through the Radio Galaxy Zoo program \citep{Banfield_2015_MNRAS_453_2326} to be double radio lobes apparently associated with late-type galaxies. We systematically and quantitatively analyze the HST images and estimate the probability of chance alignment for each object. From the original sample of 32 sources presented by \cite{Keel_2022_AJ_163_150}, we arrive at a high-confidence sample of 18 sources that we deem to be truly associated with the optical counterpart imaged with HST. We demonstrate that extremely late-type disk galaxies can indeed produce extended radio jets, but these galaxies turn out to be very massive and mostly red. We argue that BH mass, not only BH spin, is a key factor governing jet production.

The paper is structured as follows. Section~\ref{sec:Sample} summarizes the observations used in this paper. We evaluate the probability of chance alignment and establish a high-confidence sample in Section~\ref{sec:chance}. We describe our measurements of the radio and optical images in Section~\ref{sec:measure} and present the statistical properties of the host galaxies and radio sources in Section~\ref{sec:results}. Section~\ref{sec:discuss} discusses the connection between launching radio lobes and the physical properties of the host galaxies. This paper adopts a \cite{Chabrier_2003_ApJ} stellar initial mass function and a $\Lambda$CDM cosmology with $H_0 = 69\,\mathrm{ km\, s^{-1} \,Mpc^{-1}}$, $\Omega_\mathrm{m} = 0.29$, and $\Omega_\Lambda = 0.71$ based on the final nine-year Wilkinson Microwave Anisotropy Probe observations \citep{Hinshaw_2013_ApJS_208_19}.

\begin{deluxetable*}{cccccccccc}
\tablecaption{Basic Properties of the Sample \label{tab:basic}}
\tablewidth{0pt}
\tablehead{
\colhead{Name} &\colhead{Full Name} & \colhead{R. A. (J2000)} & \colhead{Decl. (J2000)} & \colhead{$z_\mathrm{phot}$} & \colhead{$z_\mathrm{spec}$} & \colhead{$A_V$} & \colhead{$m_g$}& \colhead{$m_r$}&\colhead{$m_i$}\\
\colhead{}& \colhead{ } & \colhead{($^{\mathrm{h}}\;{}^{\mathrm{m}}\;{}^{\mathrm{s}}$)} & \colhead{($^\circ\;{}^\prime\;{}^{\prime\prime}$)}  & \colhead{ } & \colhead{ } & \colhead{(mag)} & \colhead{(mag)} & \colhead{(mag)} & \colhead{(mag)}
}\decimalcolnumbers
\startdata
J0209+075 & SDSS J020904.75+075004.5 & 02 09 04.75 & 07 50 04.5 & $0.251\pm0.024$ & \nodata & 0.15 & 18.80 & 17.47 & 16.97 \\
J0219+015 & UGC 1797 & 02 19 58.73 & 01 55 48.7 & $0.033\pm0.016$ & 0.041 & 0.13 & 14.48 & 13.59 & 13.12 \\
J0802+115 & SDSS J080259.73+115709.7 & 08 02 59.73 & 11 57 09.7 & $0.173\pm0.023$ & \nodata & 0.07 & 18.79 & 17.92 & 17.42 \\
J0806+062 & SDSS J080658.46+062453.4 & 08 06 58.46 & 06 24 53.4 & $0.112\pm0.055$ & \nodata & 0.06 & 18.60 & 18.20 & 17.89 \\
J0813+552 & SDSS J081303.10+552050.7 & 08 13 03.10 & 55 20 50.7 & $0.279\pm0.035$ & \nodata & 0.12 & 20.39 & 19.51 & 19.13 \\
J0823+033 & SDSS J082312.91+033301.3 & 08 23 12.91 & 03 33 01.3 & $0.122\pm0.011$ & \nodata & 0.08 & 16.63 & 15.59 & 15.11 \\
J0832+184 & SDSS J083224.82+184855.4 & 08 32 24.82 & 18 48 55.4 & $0.122\pm0.005$ & 0.114 & 0.09 & 17.11 & 16.05 & 15.60 \\
J0833+045 & SDSS J083351.28+045745.4 & 08 33 51.28 & 04 57 45.4 & $0.246\pm0.03$ & \nodata & 0.09 & 19.77 & 18.99 & 18.64 \\
J0847+124 & SDSS J084759.90+124159.3 & 08 47 59.90 & 12 41 59.3 & $0.157\pm0.025$ & 0.175 & 0.07 & 18.38 & 17.15 & 16.61 \\
J0855+420 & B3 0852+422 & 08 55 49.15 & 42 04 20.1 & $0.191\pm0.025$ & \nodata & 0.08 & 19.45 & 18.14 & 17.55 \\
J0901+164 & SDSS J090147.17+164851.3 & 09 01 47.17 & 16 48 51.3 & $0.232\pm0.026$ & \nodata & 0.08 & 19.28 & 18.26 & 17.79 \\
J0903+432 & SDSS J090305.84+432820.4 & 09 03 05.84 & 43 28 20.4 & $0.369\pm0.041$ & 0.373 & 0.05 & 20.70 & 19.17 & 18.46 \\
J0914+413 & B3 0911+418 & 09 14 45.54 & 41 37 14.3 & $0.149\pm0.010$ & 0.140 & 0.05 & 16.35 & 15.22 & 14.72 \\
J0919+135 & SDSS J091949.07+135910.7 & 09 19 49.07 & 13 59 10.7 & $0.417\pm0.049$ & \nodata & 0.10 & 21.47 & 20.50 & 20.23 \\
J0926+465 & SDSS J092605.17+465233.9 & 09 26 05.17 & 46 52 33.9 & $0.202\pm0.040$ & \nodata & 0.04 & 19.46 & 18.21 & 17.74 \\
J0941+312 & B2 0938+31A & 09 41 03.63 & 31 26 18.7 & $0.366\pm0.037$ & \nodata & 0.05 & 20.93 & 19.56 & 19.00 \\
J0956+162 & SDSS J095605.87+162829.9 & 09 56 05.87 & 16 28 29.9 & $0.341\pm0.066$ & \nodata & 0.09 & 20.43 & 19.18 & 18.77 \\
J0958+561 & SDSS J095833.44+561937.8 & 09 58 33.44 & 56 19 37.8 & $0.247\pm0.021$ & \nodata & 0.03 & 19.05 & 17.78 & 17.35 \\
J1128+241 & SDSS J112811.63+241746.9 & 11 28 11.63 & 24 17 46.9 & $0.121\pm0.022$ & \nodata & 0.05 & 17.88 & 17.34 & 17.04 \\
J1136+125 & SDSS J113648.57+125239.7 & 11 36 48.57 & 12 52 39.7 & $0.059\pm0.021$ & 0.034 & 0.07 & 17.37 & 17.02 & 16.80 \\
J1303+511 & SDSS J130300.80+511954.7 & 13 03 00.80 & 51 19 54.7 & $0.122\pm0.037$ & \nodata & 0.03 & 19.09 & 18.35 & 17.95 \\
J1322+270 & IC 4234 & 13 22 59.87 & 27 06 59.1 & $0.027\pm0.005$ & 0.034 & 0.06 & 14.26 & 13.63 & 13.31 \\
J1328+571 & SDSS J132809.31+571023.3 & 13 28 09.31 & 57 10 23.3 & $0.032\pm0.029$ & \nodata & 0.03 & 16.91 & 16.78 & 16.69 \\
J1349+454 & SDSS J134900.13+454256.5 & 13 49 00.13 & 45 42 56.5 & $0.273\pm0.042$ & \nodata & 0.04 & 20.83 & 19.45 & 18.85 \\
J1354+465 & B3 1352+471 & 13 54 36.02 & 46 57 01.5 & $0.180\pm0.034$ & \nodata & 0.05 & 20.65 & 19.97 & 19.62 \\
J1509+515 & SDSS J150903.21+515247.9 & 15 09 03.21 & 51 52 47.9 & $0.575\pm0.031$ & 0.579 & 0.06 & 20.81 & 19.54 & 18.72 \\
J1633+084 & SDSS J163300.85+084736.4 & 16 33 00.85 & 08 47 36.4 & $0.266\pm0.032$ & \nodata & 0.17 & 18.96 & 17.69 & 17.20 \\
J1636+243 & SDSS J163624.97+243230.8 & 16 36 24.97 & 24 32 30.8 & $0.118\pm0.040$ & \nodata & 0.11 & 20.10 & 19.60 & 19.33 \\
J1646+383 & B2 1644+38 & 16 46 28.42 & 38 31 16.0 & $0.098\pm0.021$ & 0.108 & 0.04 & 17.51 & 16.63 & 16.17 \\
J1656+640 & SDSS J165620.60+640752.9 & 16 56 20.60 & 64 07 52.9 & $0.201\pm0.018$ & 0.212 & 0.07 & 17.98 & 16.90 & 16.40 \\
J1721+262 & SDSS J172107.89+262432.1 & 17 21 07.89 & 26 24 32.1 & $0.147\pm0.011$ & 0.170 & 0.11 & 18.78 & 17.68 & 17.18 \\
J2141+082 & SDSS J214110.61+082132.6 & 21 41 10.61 & 08 21 32.6 & $0.307\pm0.043$ & \nodata & 0.15 & 20.22 & 19.11 & 18.43 \\
\enddata
\tablecomments{ Col. (1): Abbreviated object name. Col. (2): Full object name from \cite{Keel_2022_AJ_163_150}. Cols. (3)--(4): Coordinates. Col. (5): Photometric redshift from the DESI survey. Col. (6): Spectroscopic redshift; all except object J0219+015 \citep{Huchra_1999_ApJS_121_287} derive from SDSS. Col. (7): Galactic extinction from \cite{Schlafly_2011_ApJ_737_103}. Cols. (8)--(10): SDSS $gri$-band {\tt model} magnitude after correction for Galactic extinction using the extinction curve of \cite{Cardelli_1989_ApJ_345_245}.}
\end{deluxetable*}

\begin{figure*}[]
\centering
\includegraphics[width=1\textwidth]{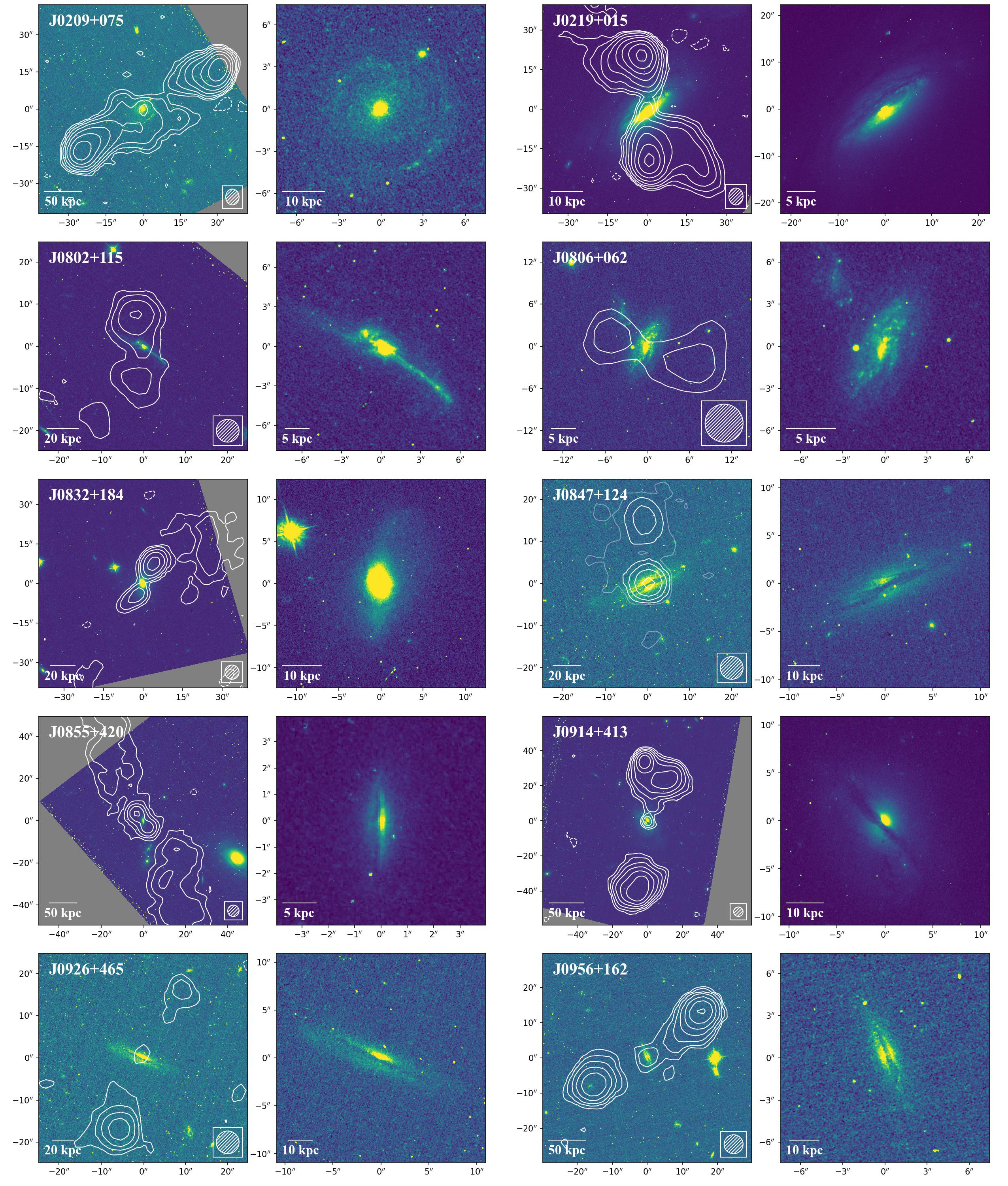}
\caption{HST F475W images of the 18 high-confidence sources in our sample, overlaid with FIRST 1.4\,GHz contours in the left panel and zoomed-in to highlight the optical morphology of the galaxy in the right panel. All images are centered on the galaxy with north up and east to the left. The restoring beam of the radio map is depicted as a hatched ellipse on the lower-right corner. Radio contours are ($-3, 3, 6, 12, 24, 48, 96, ...$) $\times$ rms of each image, where the values of the rms are listed in Table~3. Object J0847+124 shows an additional 1.5 rms contour in light-grey color. The scale bar in the lower-left corner of each panel indicates the proper distance at the redshift of each object. Figure~A1 displays the low-confidence objects.}
\label{fig:figset}
\end{figure*}

\begin{figure*}[]
\figurenum{1}
\centering
\includegraphics[width=1\textwidth]{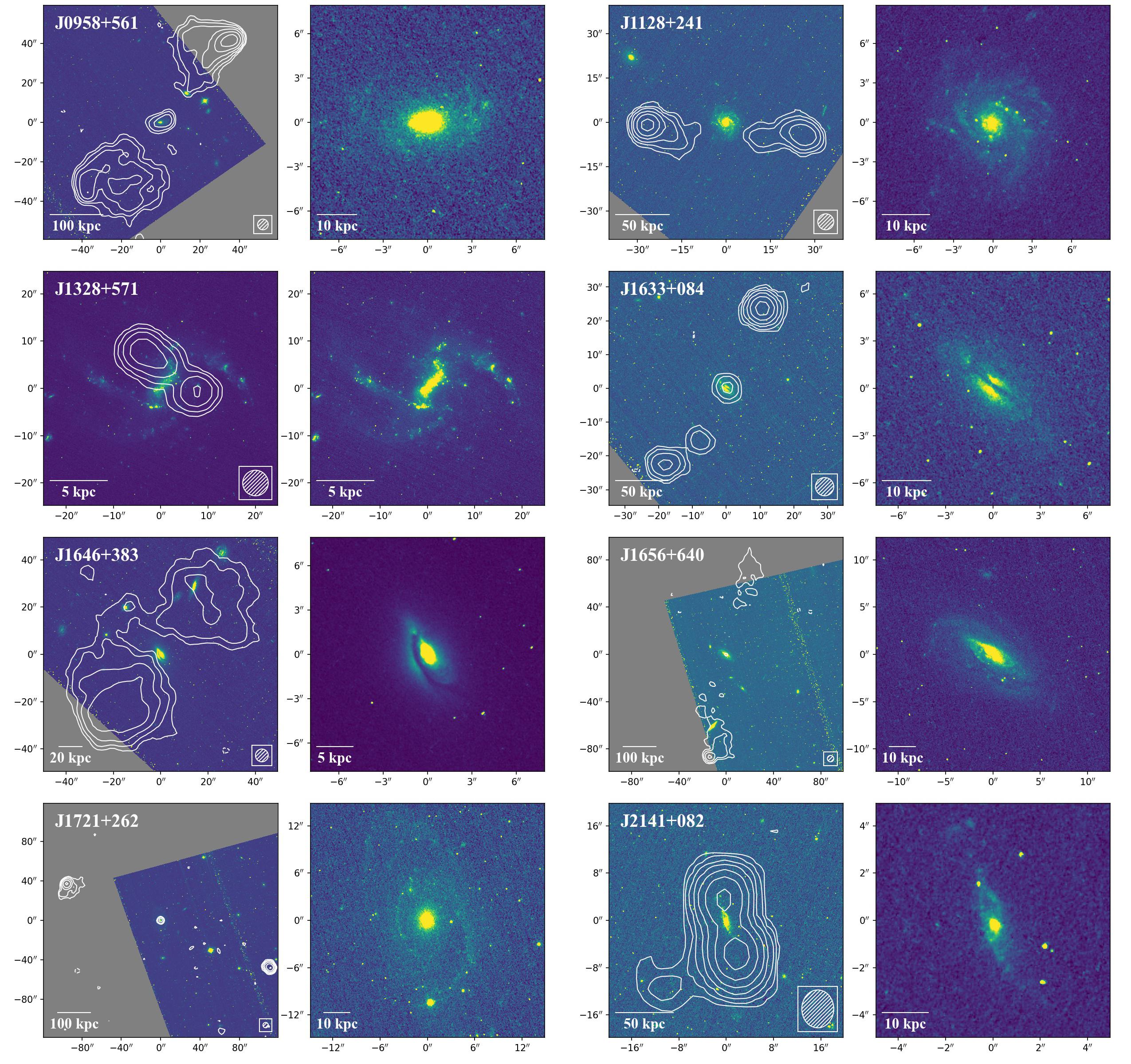}
\caption{\textit{(continued)}}
\end{figure*}

\section{Observations}
\label{sec:Sample}

Our sample is based on the Radio Galaxy Zoo and the Zoo Gems projects. Radio Galaxy Zoo is an online citizen-science program that invites volunteers to identify infrared and optical counterparts of radio sources in the Faint Images of the Radio Sky at Twenty Centimeters \citep[FIRST;][]{Becker_1995_ApJ_450_559} survey. The FIRST survey gives snapshots of the sky in the 1.4\,GHz band with an exposure time of 3 minutes, which typically generates images with a full-width at half maximum (FWHM) beam size of 5\farcs4, an astrometric accuracy of 50$\,$mas, and root mean square (rms) noise of 0.15$\,$mJy\,beam$^{-1}$. Radio Galaxy Zoo includes all spatially resolved radio sources in the FIRST survey with signal-to-noise ratios higher than 10. The radio images are fed to volunteers to locate the host galaxies by overlaying them with optical images from Data Release 10 \citep{Ahn_2014_ApJS} and Data Release 12 \citep{Alam_2015_ApJS} of the Sloan Digital Sky Survey (SDSS) and with near-infrared images from the Wide-Field Infrared Survey Explorer \citep[WISE;][]{Wright_2010_AJ_140_1868}.

Zoo Gems is a follow-up project of Radio Galaxy Zoo that makes use of short windows in the HST schedule to observe galaxies with peculiar morphologies of interest. Before the start of Zoo Gems, roughly half of the candidate FIRST sources had host identifications and were sent to the data pool of Zoo Gems for further analysis. Zoo Gems selected 32 double-lobed radio sources apparently associated with a host galaxy having a prominent disk component, according to analysis based on the SDSS pipeline and two-dimensional fits using \GALFIT \ \citep{Peng_2002_AJ_124_266, Peng_2010_AJ_139_2097}. HST observations of 674\,s duration were obtained in the F475W filter (close to the SDSS $g$ band) using the Wide-Field Camera mode of the Advanced Camera for Surveys \citep{Ford_1998}. A two-point dither pattern was adopted to better sample the point-spread function. The images have a pixel scale of $0\farcs05$, a resolution of ${\rm FWHM} \approx 0\farcs10-0\farcs14$, and an astrometry accuracy of $\lesssim 10\,\mathrm{mas}$. Using these 32 objects as our initial sample, we retrieved the radio images from the FIRST archive\footnote{http://sundog.stsci.edu/cgi-bin/searchfirst} and the optical HST images from the Mikulski Archive for Space Telescopes (MAST)\footnote{https://mast.stsci.edu/search/hst/ui/}. Figure~\ref{fig:figset} displays the HST images overlaid with the FIRST radio contours, for the subset of 18 sources we regard as having genuine optical-radio association (see details in Section~3); the images for the objects with less confident association are presented in Appendix~A.

All galaxies have SDSS multiband photometry, and for our purposes we collect their {\tt model} magnitudes in the $g$, $r$, and $i$ band (Table~\ref{tab:basic}). Only 11 objects have spectroscopic redshifts ($z_{\rm spec}$). For the remaining 21 objects, we adopt photometric redshifts ($z_{\rm phot}$) from \cite{Duncan_2022_MNRAS_512_3662}, which is based on the Dark Energy Spectroscopic Instrument (DESI) Legacy Imaging Surveys \citep{Dey_2019_AJ_157_168}. DESI covers most of the footprint of SDSS in $grz$, with $5\,\sigma$ point source depths reaching $m_g=24.7$, $m_r=23.9$, $m_z=23.0$\,mag at $1\farcs5$ seeing, $\sim 1.5$\,mag deeper than SDSS. \cite{Duncan_2022_MNRAS_512_3662} derived photometric redshifts using a machine-learning technique, which is based on galaxy properties in the optical and mid-infrared bands, including color, magnitude, and size. Their photometric redshifts have a robust scatter compared with the spectroscopic redshift, with a normalized median absolute deviation $1.48 \times \mathrm{median}(|z_{\mathrm{phot}}- z_{\mathrm{spec}} |/\left(1+z_{\mathrm{spec}})\right) = 0.05$. Comparing $z_{\rm phot}$ with $z_{\rm spec}$ for the 11 objects with spectra yields good consistency with a scatter of only 0.01 for (1+$z$). 

For completeness, we note that there is a mismatch between the target names and sky coordinates in Table~2 of \cite{Keel_2022_AJ_163_150}. The names always lead to correct targets, as judged by the ancillary information available in NASA/IPAC Extragalactic Database (NED). Some sources are named by radio programs (e.g., B3~0911+418), and their NED coordinates pertain to the center position of the radio emission instead of the optical center of the host galaxy. For consistency, all the updated coordinates in Table~\ref{tab:basic} refer to the position of the optical center of the galaxy.

\section{Sample Definition}
\label{sec:chance}

The original sample of double-lobed radio sources from the Zoo Gems project was defined by their apparent association with an optically visible disk galaxy. To be fully convinced that the radio-optical association is real, we assess the probability that the galaxies we observe are aligned fortuitously with the radio lobes. The identification of the hosts of radio sources is usually based on the relative position between the radio emission and optically visible galaxies. However, because radio lobes lack emission lines to determine their redshift, the line-of-sight distance between the radio lobes and galaxies is unknown. Other sources along the line-of-sight might happen to be projected toward the radio center and be mistaken for the host galaxy. To statistically rule out such coincidences, we calculate the probability that the observed optical counterparts are chance-aligned, while the true host galaxies (assumed to be elliptical galaxies) are too faint to be detected. We use this chance alignment probability to define the final sample of galaxies included in this paper.

\subsection{Probability of Disk Galaxies in Chance Alignment} \label{sec:chance late}
We calculate the probability that an unrelated disk galaxy happens to be projected toward the center of the radio lobes. To begin with, the chance alignment probability depends on the number density of such galaxies. We use the \cite{Schechter_1976_ApJ} luminosity function to calculate the number density of disk galaxies, which, in terms of magnitude, is \citep{Loveday_2012_MNRAS_420_1239}

\small
\begin{equation}
\phi(M) =0.4 \ln 10\;\phi^{*}\left\{10^{0.4\left(M^{*}-M\right)}\right\}^{\alpha+1} \exp \left\{-10^{0.4\left(M^{*}-M\right)}\right\},
\end{equation}
\normalsize

\noindent
where $M^*$ is the characteristic absolute magnitude, $\alpha$ is the power index, and $\phi ^{*}$ is the normalization of the number density in units of $\mathrm{\;Mpc^{-3}}$. We adopt the $g$-band luminosity function of blue galaxies from \cite{Loveday_2012_MNRAS_420_1239}, for which $M^* - 5\log\, h = -19.6$, $\phi^* = 7.1 \times 10^{-3} \,h^{-3}\mathrm{\,Mpc^{-3}}$, and $\alpha = -1.42$, with $h$ the dimensionless Hubble constant. For galaxies brighter than a given apparent magnitude $m_0$, we calculate the surface number density $n(m \le m_0)$ by integrating over all redshift range in a unit solid angle, taking into account K-correction with a color $g-r=0.66\,$mag, typical of Sab galaxies \citep{Fukugita_1995_PASP_107_945}. 

Some galaxies have a radio core in their nucleus, but their number density is even smaller if their radio-loudness is high. Studying the radio-to-optical flux ratio 

\begin{equation}
R_B = f_\mathrm{1.4}/(k\cdot 10^{-0.4m_B}),
\label{eq:loundness}
\end{equation}

\noindent
where $f_\mathrm{1.4}$ is the 1.4\,GHz flux density in units of mJy, $m_B$ is the $B$-band magnitude of the galaxy, and $k = 4.44 \times 10^6$ is a normalization factor, \cite{Gavazzi_1999_AAP_343_86} found that only $\sim 0.4\%$ of late-type galaxies have $R_B\ge 10$ and $\sim 4\%$ have $R_B \approx 1-10$. We derive the galaxy $B$-band magnitude from SDSS $g$ and $r$ magnitudes following \cite{Jester_2005_AJ_130_873}, after considering K-correction and Galactic extinction. We measure the nuclear radio emission with the Common Astronomy Software Application ({\texttt{CASA}}; \citealt{McMullin_2007}) and apply K-correction assuming a spectral index of $-0.8$ \citep{Blundell_1999_AJ_117_677} to derive the radio luminosity. Note that here we assume that the associations with the radio lobes are unknown and only take the core radio emission into consideration. Depending on the radio-loudness, we multiply the rarity factors (0.004 for $R_B\ge 10$ and 0.04 for $R_B = 1-10$) to their number density.

Finally, following \cite{Bloom_2002_AJ_123_1111} and \cite{Berger_2010_ApJ_722_1946}, we calculate the probability that a galaxy coincides with the radio center within a separation $r_\mathrm{offset}$:

\begin{equation}
p= 1- e^{-\pi r_\mathrm{offset}^2 n(m\le m_\mathrm{gal})}\approx \pi r_\mathrm{offset}^2 n(m\le m_\mathrm{gal}),
\end{equation}

\noindent
where $n(m\le m_\mathrm{gal})$ in units of arcsec$^{-2}$ is the surface number density of disk galaxies of the same or brighter magnitude, and $r_\mathrm{offset}$ is in units of arcsec. As in \cite{Laing_1983_MNRAS_204_151}, the radio center is defined as the midpoint of the two radio hotspots.

\subsection{Probability of Observing No Elliptical Hosts} \label{sec:chance ellip}
If the disk galaxies we observe are not the hosts of the radio lobes, the real hosts, which are usually giant elliptical galaxies, should appear in the deep HST images, unless they are too faint. Therefore, we estimate the probability distribution of the magnitude of possible elliptical hosts.

\subsubsection{Probability Distribution of Host Magnitude}
\begin{figure*}[]
\includegraphics[width=1\textwidth]{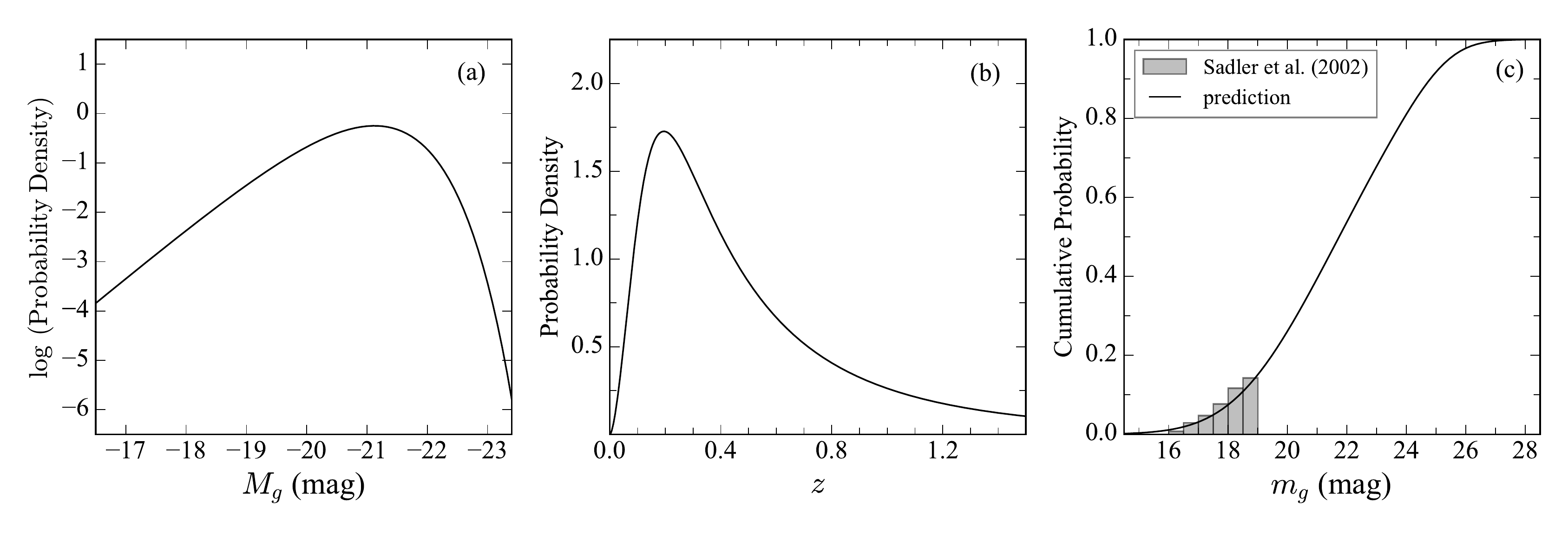}
\caption{(a) Probability density as a function of $g$-band absolute magnitude for red galaxies from \cite{Loveday_2012_MNRAS_420_1239}. (b) Predicted probability density as a function of redshift for galaxies with radio lobes and $S_\mathrm{tot}=50\,\mathrm{mJy}$. (c) Cumulative probability distribution of a potential elliptical host of radio lobes as a function of $g$-band apparent magnitude. Histograms denote objects with detected radio lobes and $S_\mathrm{tot} = 30-70\,\mathrm{mJy}$ \citep{Sadler_2002_MNRAS_329_227}. \label{fig: faintellip}}
\end{figure*}

The probability distribution of luminosity for a galaxy is proportional to its luminosity function. \cite{Scarpa_2001_ApJ_556_749} found that the luminosity function of radio elliptical galaxies correlates with that of normal elliptical galaxies as $A \phi(L)L^2$, with $L$ the galaxy luminosity, $\phi(L)$ the luminosity function for normal elliptical galaxies, and a normalization factor $A$. We adopt $\phi(L)$ of red galaxies from \cite{Loveday_2012_MNRAS_420_1239} and derive the probability density distribution of absolute magnitude for radio elliptical galaxies (Figure~\ref{fig: faintellip}a). Our order-of-magnitude estimates, which suffice for our purposes, do not consider evolutionary effects of the luminosity function. Section~\ref{sec:chance_total} will show that this assumption provides a conservative estimate of chance-alignment probability. 

Meanwhile, we estimate the probability distribution of redshift based on the radio luminosity function and the observed radio flux density (Section~\ref{sec:measure}). We describe the radio luminosity function as a double power law, 

\begin{equation}
\Phi(L_\mathrm{1.4})=\frac{\Phi^*}{\left(L_\mathrm{1.4}/ L^*\right)^{\alpha}+\left(L_\mathrm{1.4}/ L^*\right)^{\beta}},
\end{equation}

\noindent
where $L_\mathrm{1.4}$ is the 1.4\,GHz luminosity in units of $\mathrm{W} \mathrm{\,Hz}^{-1}$, $L^*$ is the characteristic radio luminosity, and $\Phi^*$ is the characteristic number density. Following \cite{Mauch_2007_MNRAS_375_931}, we adopt $\Phi^*=10^{-5.5} \,\mathrm{mag}^{-1}\, \mathrm{Mpc}^{-3}$, $L^*=10^{24.6} \,\mathrm{W \, Hz}^{-1}$, $\alpha=1.27 $, and $\beta=0.49$. The probability that a radio source with a flux density $S$ is at redshift $z$ is proportional to the number of sources at that redshift:

\begin{equation}
p(z,S)\,dz\,dS=\frac{1}{\kappa}\,\Phi (L_\mathrm{1.4})\, d\log\, L_\mathrm{1.4}\, dV,
\end{equation}

\noindent
where $\kappa$ is a normalization factor, $L_\mathrm{1.4}$ and $V$ are the corresponding absolute luminosity and comoving volume at redshift $z$, and $L_\mathrm{1.4}= S\cdot 4\pi d_L/(1+z)^{1+\alpha}$, with $d_L$ the luminosity distance and $\alpha=-0.8$ the spectral index \citep{Blundell_1999_AJ_117_677}. Figure~\ref{fig: faintellip}b shows the probability density as a function of redshift, for a typical radio flux density in our sample, $S_\mathrm{tot}=50\,$mJy (Table~\ref{tab:radio}). The probability peaks at $z\approx0.2$ and decreases rapidly toward both lower and higher redshift.

Finally, we calculate the probability distribution of the apparent magnitude $m$ by combining those of the absolute magnitude and the redshift, considering 

\begin{equation}
m=M+\mu(z)+K(z),
\end{equation}

\noindent
where $\mu(z)$ is the distance modulus as a function of redshift, defined as $\mu = 5 \log\,(d_L/10\,\rm{pc})$, and $K(z)$ is the K-correction for elliptical galaxies, for which we use the spectral energy distribution (SED) templates generated using a delayed star formation history with an e-folding time of 2.5$\,$Gyr. This set of templates is based on optical and near-infrared SEDs of elliptical and S0 galaxies in the Carnegie-Irvine Galaxy Survey \citep[CGS;][]{Ho_2011_ApJS_197_21}, which match well the typical star formation history of early-type galaxies from the GAMA survey \citep{Bellstedt+2020MNRAS}. We compare our probability distribution (Figure~\ref{fig: faintellip}c) with a large sample of galaxies from the complete radio survey\footnote{We transform their magnitudes in the $b_J$ band to the $g$ band assuming $b_J - g = 0.5\,\rm{mag}$ \citep{Jester_2005_AJ_130_873,Chang_2006_MNRAS_366_717}.} of \cite{Sadler_2002_MNRAS_329_227}. We find excellent agreement between the observations and our model if we constrain the radio flux density to the same range ($S_\mathrm{tot} = 30-70\,$mJy). Moreover, we note that the probability distribution is insensitive to the value of the observed radio flux density in the range of this sample ($6-374\,$mJy). For instance, if $m_g < 25\,$mag, the typical flux density (50\,mJy) gives a probability $\sim 0.1$, while the faintest case (6\,mJy) gives a probability of $\sim 0.15$. Since we only need an order-of-magnitude estimate of the probability, we adopt the typical results for all objects.

\subsubsection{Source Detection and Constraint on Host Magnitude}
Might the hosts of the radio sources actually be elliptical galaxies that have simply escaped detection? We use the {\tt Python} source detection package \Phot \ \citep{larry_bradley_2020_4044744} to search a region centered at the midpoint of the radio hotspots. Defining $d$ as the angular distance between the two hotspots, \cite{Laing_1983_MNRAS_204_151} find that 96\% of optical counterparts are distributed within $0.2\,d$ from the radio center. We thus restrict our search to a radius of $0.2\,d$. We detect sources with a threshold set to 3 times the rms noise per pixel, requiring that the source contains 40 connected (0.1$\,$arcsec$^2$) pixels, which is one-tenth of the minimal area of a galaxy of radius 5$\,$kpc across the entire redshift range (the angular-diameter scale peaks at $\mathrm{8.6\,kpc\,arcsec^{-1}}$ in our cosmology). We further crossmatch the detected sources with the DESI Legacy Imaging Surveys Data Release 8 \citep{Duncan_2022_MNRAS_512_3662} to obtain their redshift and exclude local small galaxies from candidate giant ellipticals, according to their absolute magnitude and size. Appendix \ref{sec:appendix} describes four positive detections of possible candidates. To be conservative, we view them as candidate hosts. For the other cases, we constrain the hypothetical elliptical galaxy host to be fainter than our detection limit, and we estimate the probability of it being so faint.

To estimate the detection limits, we randomly add simulated elliptical galaxies to the HST images and perform source detection as described above to determine the detection rate, which is defined as the percentage of detected cases. We use {\tt GALFIT} to simulate single-component \cite{Sersic_1968} models with index $n=4$, as typically seen in elliptical galaxies \citep{deVaucouleurs_1948}. We set the effective radii to $r_e = 3\farcs0$, $1\farcs6$, and $1\farcs2$, which correspond to typical radio galaxies with $r_e \approx 10\,$kpc \citep{McLure_1999_MNRAS_308} at $z= 0.2$, 0.5, and 1.2. To examine our capability to detect sources with $m_g = 27\,$mag, we simulate 100 mock elliptical galaxies for each $r_e$. All the HST images have detection rates higher than 97\% for each value of $r_e$, which suggests that the detection limit is $\gtrsim 27\,$mag. Nevertheless, we do not adopt $m_g = 27\,$mag as the detection limit in our analysis, for too many random sources ($> 1$) would fall into our detection regions with this magnitude limit. Instead, we only regard sources brighter than 25$\,$mag as candidate hosts. This much more conservative limit suffices for our chance-alignment probability analysis and minimizes the number of confusing objects ($\lesssim 0.3$) in our detection regions. We additionally consider the fact that our detection would be less sensitive if the sources overlap with a foreground disk galaxy, the likelihood of which depends on the size of the disk galaxy and the spatial distribution of the latent hosts. While 96\% of optical counterparts are distributed within $0.2\,d$ from the radio center \citep{Laing_1983_MNRAS_204_151}, the chance of overlap is $\sim 50\%$ when a disk galaxy with $r_e = 0.1\,d$ is located in the center. Therefore, the chance of overlap is significant for disk galaxies with $r_e>0.1 \,d$.

In the case of blended sources, we estimate the detection limit by adding simulated elliptical galaxies of increasing brightness to the region within the source's effective radius, until reaching the critical magnitude beyond which they are hardly discernible through visual inspection of the residual images from \GALFIT \ decomposition (Section~4.2). Our tests indicate a limiting magnitude of $m_g \approx 22$. The probability of an elliptical host being fainter than the limit is 0.4, according to Figure~\ref{fig: faintellip}c. Incidentally, the specific value of the detection limit does not impact substantially our probability analysis for limiting magnitudes in the range $m_g \approx 20-24$ (Figure~\ref{fig: faintellip}c). Although our estimate of the detection limit is approximate, it is sufficient for our analysis. In the end, we conclude that the non-detection probability is $\sim 0.4$ when the chances of overlapping are high, and $\sim 0.1$ when the chances are low.

\begin{deluxetable*}{ccchccccc}
\tablecaption{Sample Definition Based on Probability of Chance Alignment \label{tab:chance}}
\tablewidth{0pt}
\tablehead{
\colhead{Name} &\colhead{$r_\mathrm{offset}$} & \colhead{$R_B$} &\nocolhead{magnitude} & \colhead{Surface Density} & \colhead{Overlap} & \colhead{$P$} &\colhead{$N_c$}\\
\colhead{}& \colhead{(\arcsec)} & \colhead{ }& \nocolhead{(mag)} & \colhead{(deg$^{-2}$)} & \colhead{ } & \colhead{}&\colhead{} \\
\colhead{(1)} & \colhead{(2)} & \colhead{(3)} & \nocolhead{}&\colhead{(4)} & \colhead{(5)} & \colhead{(6)} & \colhead{(7)}
}
\startdata
\\
\multicolumn{3}{l}{{\it High-confidence Sample}} &&&&&\\
J0847+124 & 0.7 & 23.8 & 18.4 & 24 & H & $4.6 \times 10^{-9}$ & 0.0002 \\
J0219+015 & 1.1 & \nodata & 14.5 & 0.17 & H & $2.0 \times 10^{-8}$ & 0.0010 \\
J0914+413 & 5.4 & 3.9 & 16.4 & 2.06 & L & $5.8 \times 10^{-8}$ & 0.0029 \\
J1633+084 & 3.7 & 21.5 & 19.0 & 49 & L & $6.6 \times 10^{-8}$ & 0.0033 \\
J1646+383 & 0.9 & \nodata & 17.5 & 8.1 & L & $1.6 \times 10^{-7}$ & 0.0080 \\
J0956+162 & 2.7 & 56.1 & 20.4 & 248 & L & $1.7 \times 10^{-7}$ & 0.0087 \\
J0958+561 & 7.5 & 43.9 & 19.1 & 56 & L & $3.0 \times 10^{-7}$ & 0.015 \\
J1721+262 & 9.3 & 18.2 & 18.8 & 39 & L & $3.3 \times 10^{-7}$ & 0.016 \\
J0832+184 & 1.4 & \nodata & 17.1 & 5.0 & H & $9.4 \times 10^{-7}$ & 0.047 \\
J0926+465 & 1.7 & 5.1 & 19.5 & 89 & H & $1.0 \times 10^{-6}$ & 0.050 \\
J1656+640 & 9.5 & $\lesssim$2 & 18.0 & 15 & L & $1.3 \times 10^{-6}$ & 0.065 \\
J0806+062 & 0.7 & \nodata & 18.6 & 31 & H & $1.5 \times 10^{-6}$ & 0.073 \\
J0855+420 & 0.5 & \nodata & 19.4 & 79 & H & $1.9 \times 10^{-6}$ & 0.096 \\
J1128+241 & 2.6 & \nodata & 17.9 & 13 & L & $2.2 \times 10^{-6}$ & 0.11 \\
J2141+082 & 1.1 & \nodata & 20.2 & 198 & L & $5.8 \times 10^{-6}$ & 0.29 \\
J1328+571 & 4.0 & \nodata & 16.9 & 3.9 & H & $6.0 \times 10^{-6}$ & 0.30 \\
J0209+075 & 2.7 & \nodata & 18.8 & 39 & L & $6.9 \times 10^{-6}$ & 0.34 \\
J0802+115 & 1.5 & \nodata & 18.8 & 39 & H & $8.5 \times 10^{-6}$ & 0.42 \\
\\
\multicolumn{3}{l}{{\it Low-confidence Sample}} &&&&&\\
J0833+045 & 2.4 & \nodata & 19.8 & 126 & L & $1.8 \times 10^{-5}$ & 0.88 \\
J1354+465 & 2.3 & \nodata & 20.7 & 345 & L & $4.4 \times 10^{-5}$ & 2.2 \\
J1636+243 & 3.8 & \nodata & 20.1 & 177 & L & $6.2 \times 10^{-5}$ & 3.1 \\
J0901+164 & 2.0 & \nodata & 19.3 & 70 & L & $6.8 \times 10^{-5}$ & 3.4 \\
J1136+125 & 7.2 & \nodata & 17.4 & 7.2 & L & $9.0 \times 10^{-5}$ & 4.5 \\
J1509+515 & 3.8 & \nodata & 20.8 & 384 & L & $1.3 \times 10^{-4}$ & 6.7 \\
J0813+552 & 1.6 & \nodata & 20.4 & 248 & L & $1.5 \times 10^{-4}$ & 7.7 \\
J0903+432 & 4.4 & \nodata & 20.7 & 345 & L & $1.6 \times 10^{-4}$ & 8.1 \\
J0919+135 & 18.1 & $\lesssim$32 & 21.5 & 804 & L & $2.6 \times 10^{-4}$ & 13 \\
J0941+312 & 8.2 & \nodata & 20.9 & 428 & L & $7.0 \times 10^{-4}$ & 35 \\
J0823+033 & \nodata & \nodata & \nodata & \nodata & \nodata & \nodata & \nodata \\
J1322+270 & \nodata & \nodata & \nodata & \nodata & \nodata & \nodata & \nodata \\
J1303+511 & \nodata & \nodata & \nodata & \nodata & \nodata & \nodata & \nodata \\
J1349+454 & \nodata & \nodata & \nodata & \nodata & \nodata & \nodata & \nodata \\
\enddata
\tablecomments{Col. (1): Object name. Col. (2): Offset between the galaxy center and the midpoint of the radio lobes. Col. (3): Radio-loudness of the radio emission in the nuclear region, defined by Equation~\ref{eq:loundness}. Here we assume that the association with the radio lobes is unknown. Col. (4): Surface number density of galaxies of the same or brighter magnitude. Col. (5): Overlapping chance that a latent elliptical host galaxy hides behind the foreground galaxy: ``L'' = low chance of overlap, corresponding to a 0.1 probability of non-detection; ``H'' = high chance of overlap, corresponding to a 0.4 probability of non-detection. Col. (6): Chance alignment probability for a single case provided the radio-optical relative location and luminosity. Col. (7): Expected number of chance-aligned cases if selecting with criteria as in this case in the entire FIRST survey, which has $N \approx 5 \times 10^4$ pairs of double-lobed sources \citep{McMahon_2002_ApJS_143_1}; $N_c = P \times N$.}
\end{deluxetable*}

\subsection{Total Probability of Chance Alignment}
\label{sec:chance_total}
We calculate the total probability of chance alignment $P$ by multiplying the probability that a disk galaxy coincides with the radio center (Section~\ref{sec:chance late}) with the probability that an elliptical host galaxy is potentially present but too faint to be detected (Section~\ref{sec:chance ellip}). \cite{McMahon_2002_ApJS_143_1} detected a total of $N\approx 5\times 10^4$ double-lobed sources in the FIRST survey. The expected number of chance-aligned events with properties specified for each object is therefore $N_c = P \times N$ (Table~\ref{tab:chance}). We compile objects with $N_c < 0.5$ into the high-confidence sample and restrict our present analysis only to these. Assuming a Poisson distribution, the high-confidence sample has 90\% probability of including no more than one dubious case. We note that there may be a few cases having real radio associations even in the low-confidence sample, as we have not taken into consideration other evidence such as AGN activity and jet orientation. 

We conclude this section with a discussion of the evolutionary effects of the luminosity function on our probability  estimation. Our results, which are based on a non-evolving luminosity function, are roughly consistent with those in an evolutionary framework because the radio sources are most likely located at low redshifts (Figure \ref{fig: faintellip}b). For a more quantitative estimation, we consider the evolving radio and optical luminosity functions of \cite{Ceraj_2018_A&A} and \cite{Loveday_2012_MNRAS_420_1239}, respectively. We find that the chance-alignment probabilities decrease by $\sim$ 70\%, which suggests that our previous estimation is very conservative. However, since the evolution parameters have large uncertainties, whose precise determination remains controversial \citep{Loveday_2012_MNRAS_420_1239, Loveday_2015_MNRAS_451_1540, Prescott_2016_MNRAS_457_730, Ocran_2021_MNRAS}, we conservatively choose to adopt the non-evolutionary results in our analysis.

\section{Data Analysis}
\label{sec:measure}
\subsection{Properties of the Radio Core and Lobes}
We use \texttt{CASA} to measure the basic radio properties of the sources in the FIRST images (Table~\ref{tab:radio}). We determine the rms noise from the standard deviation of pixel values in the background regions far from the main source. We measure the integrated flux density of the radio lobes and radio core, if present, with polygons that enclose all the radio emission. Provided the redshift of the host galaxy, we infer the absolute radio luminosity $L_{1.4}$ after K-correction assuming a spectral index of $-0.8$ \citep{Blundell_1999_AJ_117_677}. We fit the hotspots and radio core with a two-dimensional Gaussian profile, respectively, to measure their positions and flux density. We determine the separation between the two hotspots, their mean position angle (PA) relative to the galactic center, and the projected proper distance of the hotspots. We visually classify the radio sources into two types following the precepts of \cite{Fanaroff_1974_MNRAS_167_31P}, according to whether the distance between the two hotspots is smaller than half of the total extent of the double lobes (FR~I) or not (FR~II).

\begin{deluxetable*}{ccccccCCCCCCC}[b]
\tablecaption{ Measurement of Radio Properties\label{tab:radio}}
\tablewidth{0pt}
\tablehead{
\colhead{Name} & \colhead{Beam} & \colhead{rms} & \colhead{FR Type} &  \colhead{$d$} & \colhead{$D$} & \colhead{PA} & \colhead{$S_\mathrm{core}$} & \colhead{$S_\mathrm{tot}$} & \colhead{$\log \, L_\mathrm{1.4}$} \\
\colhead{} & \colhead{(\arcsec \ $\times$ \arcsec)} & \colhead{(mJy\,beam$^{-1}$)} & \colhead{}  & \colhead{(\arcsec)} & \colhead{(kpc)} & \colhead{($^\circ$)} & \colhead{(mJy)} & \colhead{(mJy)} & \colhead{($\mathrm{W\,Hz^{-1}}$)}\\
\colhead{(1)} & \colhead{(2)} & \colhead{(3)} & \colhead{(4)} & \colhead{(5)} & \colhead{(6)} & \colhead{(7)} & \colhead{(8)} & \colhead{(9)} & \colhead{(10)} 
}
\startdata
J0209+075 & $6.4\times5.4$ & 0.12 & II & 63 $\pm$ 4 & $250\pm23$ & 121 $\pm$ 4 & \nodata & $223 \pm 22$ & $25.62\pm0.10$ \\
J0219+015 & $6.4\times5.4$ & 0.13 & II & 39 $\pm$ 4 & $31\pm3$ & 5 $\pm$ 6 & \nodata & $315 \pm 31$ & $24.10\pm0.04$ \\
J0802+115 & $5.4\times5.4$ & 0.19 & II & 14 $\pm$ 5 & $41\pm15$ & 3 $\pm$ 23 & \nodata & $21 \pm 4$ & $24.24\pm0.14$ \\
J0806+062 & $5.4\times5.4$ & 0.17 & II & 12 $\pm$ 4 & $24\pm12$ & 73 $\pm$ 17 & \nodata & $5.7 \pm 2.9$ & $23.27\pm0.41$ \\
J0813+552 & $5.4\times5.4$ & 0.16 & II & 34 $\pm$ 4 & $145\pm21$ & 54 $\pm$ 7 & \nodata & $30 \pm 5$ & $24.85\pm0.13$ \\
J0823+033 & $6.4\times5.4$ & 0.16 & II & \nodata & \nodata & \nodata & $16.3 \pm 0.5$ & $147 \pm 15$ & $24.76\pm0.09$ \\
J0832+184 & $5.4\times5.4$ & 0.11 & II & 14 $\pm$ 4 & $29\pm8$ & 151 $\pm$ 15 & \nodata & $26 \pm 4$ & $23.94\pm0.06$ \\
J0833+045 & $5.4\times5.4$ & 0.15 & II & 26 $\pm$ 10 & $101\pm40$ & 21 $\pm$ 23 & \nodata & $25 \pm 5$ & $24.65\pm0.14$ \\
J0847+124 & $5.4\times5.4$ & 0.16 & II & 28 $\pm$ 3 & $84\pm9$ & 4 $\pm$ 6 & $5.1 \pm 0.4$ & $7.8 \pm 3.3$ & $23.82\pm0.15$ \\
J0855+420 & $5.4\times5.4$ & 0.14 & I & 8.4 $\pm$ 2 & $27\pm6$ & 38 $\pm$ 13 & \nodata & $100 \pm 12$ & $25.01\pm0.13$ \\
J0901+164 & $5.4\times5.4$ & 0.14 & II & 43 $\pm$ 1 & $160\pm13$ & 96 $\pm$ 2 & \nodata & $118 \pm 12$ & $25.27\pm0.11$ \\
J0903+432 & $5.4\times5.4$ & 0.14 & II & 44 $\pm$ 4 & $229\pm20$ & 116 $\pm$ 5 & \nodata & $7 \pm 4$ & $24.51\pm0.20$ \\
J0914+413 & $5.4\times5.4$ & 0.18 & II & 68 $\pm$ 10 & $169\pm24$ & 172 $\pm$ 8 & $4.7 \pm 0.6$ & $374 \pm 38$ & $25.29\pm0.04$ \\
J0919+135 & $5.4\times5.4$ & 0.20 & II & 142 $\pm$ 7 & $793\pm65$ & 160 $\pm$ 3 & $1.8 \pm 0.6$ & $233 \pm 25$ & $26.14\pm0.12$ \\
J0926+465 & $5.4\times5.4$ & 0.14 & II & 36 $\pm$ 2 & $121\pm19$ & 155 $\pm$ 3 & $0.5 \pm 0.2$ & $11 \pm 3$ & $24.11\pm0.20$ \\
J0941+312 & $5.4\times5.4$ & 0.15 & II & 151 $\pm$ 7 & $777\pm60$ & 53 $\pm$ 3 & \nodata & $143 \pm 16$ & $25.80\pm0.11$ \\
J0956+162 & $5.4\times5.4$ & 0.15 & II & 37 $\pm$ 2 & $182\pm23$ & 124 $\pm$ 3 & $5.5 \pm 1.0$ & $74 \pm 8$ & $25.44\pm0.18$ \\
J0958+561 & $5.4\times5.4$ & 0.14 & II & 92 $\pm$ 13 & $360\pm55$ & 140 $\pm$ 6 & $6.1 \pm 0.5$ & $186 \pm 19$ & $25.53\pm0.09$ \\
J1128+241 & $5.4\times5.4$ & 0.18 & II & 53 $\pm$ 3 & $116\pm19$ & 86 $\pm$ 3 & \nodata & $63 \pm 8$ & $24.38\pm0.16$ \\
J1136+125 & $5.4\times5.4$ & 0.15 & II & 187 $\pm$ 7 & $127\pm4$ & 94 $\pm$ 2 & $2.6 \pm 0.4$ & $40 \pm 7$ & $23.03\pm0.07$ \\
J1303+511 & $5.4\times5.4$ & 0.22 & II & \nodata & \nodata & \nodata & \nodata & $174 \pm 18$ & $24.83\pm0.25$ \\
J1322+270 & $5.4\times5.4$ & 0.16 & II & \nodata & \nodata & \nodata & $2.9 \pm 0.3$ & $2.9 \pm 0.3$ & $21.89\pm0.04$ \\
J1328+571 & $5.4\times5.4$ & 0.16 & II & 13 $\pm$ 3 & $8\pm7$ & 52 $\pm$ 12 & \nodata & $17 \pm 4$ & $22.61\pm0.58$ \\
J1349+454 & $5.4\times5.4$ & 0.15 & II & \nodata & \nodata & \nodata & \nodata & $6.7 \pm 0.7$ & $24.18\pm0.15$ \\
J1354+465 & $5.4\times5.4$ & 0.16 & II & 36 $\pm$ 3 & $110\pm18$ & 94 $\pm$ 5 & \nodata & $96 \pm 12$ & $24.94\pm0.17$ \\
J1509+515 & $5.4\times5.4$ & 0.17 & II & 23 $\pm$ 7 & $153\pm46$ & 140 $\pm$ 17 & \nodata & $162 \pm 16$ & $26.32\pm0.04$ \\
J1633+084 & $5.4\times5.4$ & 0.16 & II & 55 $\pm$ 2 & $227\pm20$ & 148 $\pm$ 2 & $2.8 \pm 0.4$ & $36 \pm 6$ & $24.89\pm0.13$ \\
J1636+243 & $5.4\times5.4$ & 0.15 & II & 76 $\pm$ 3 & $163\pm46$ & 39 $\pm$ 2 & \nodata & $61 \pm 8$ & $24.34\pm0.28$ \\
J1646+383 & $5.4\times5.4$ & 0.16 & II & 57 $\pm$ 11 & $113\pm21$ & 140 $\pm$ 15 & \nodata & $195 \pm 21$ & $24.77\pm0.04$ \\
J1656+640 & $5.4\times5.4$ & 0.13 & II & 150 $\pm$ 17 & $524\pm59$ & 169 $\pm$ 7 & $0.7 \pm 0.2$ & $49 \pm 9$ & $24.80\pm0.07$ \\
J1721+262 & $5.4\times5.4$ & 0.16 & II & 224 $\pm$ 3 & $656\pm8$ & 68 $\pm$ 1 & $2.5 \pm 0.6$ & $68 \pm 10$ & $24.73\pm0.06$ \\
J2141+082 & $6.4\times5.4$ & 0.15 & II & 9.6 $\pm$ 1.0 & $44\pm6$ & 13 $\pm$ 7 & \nodata & $86 \pm 9$ & $25.41\pm0.14$ \\
\enddata
\tablecomments{Col. (1): Object name. Col. (2): Beam size (major $\times$ minor axis). Col. (3): Background noise. Col. (4): FR type. Col. (5): Angular distance between the two radio hotspots. Col. (6): Projected proper distance between the two radio hotspots. Col. (7): Position angle of the hotspots relative to the galaxy center, averaged over the hotspot on each side. Col. (8): Flux density of the radio core, if any, measured from two-dimensional Gaussian fits. Col. (9): Integrated flux density of the entire radio source. Col. (10): Absolute luminosity of the entire radio source at 1.4\,GHz. Objects J0823+033, J1303+511, J1322+270, and J1349+454 have no well-defined double radio lobes and thus no lobe size and PA measurements.}
\end{deluxetable*}

\begin{figure*}[t]
\centering
\includegraphics[width=0.98\textwidth]{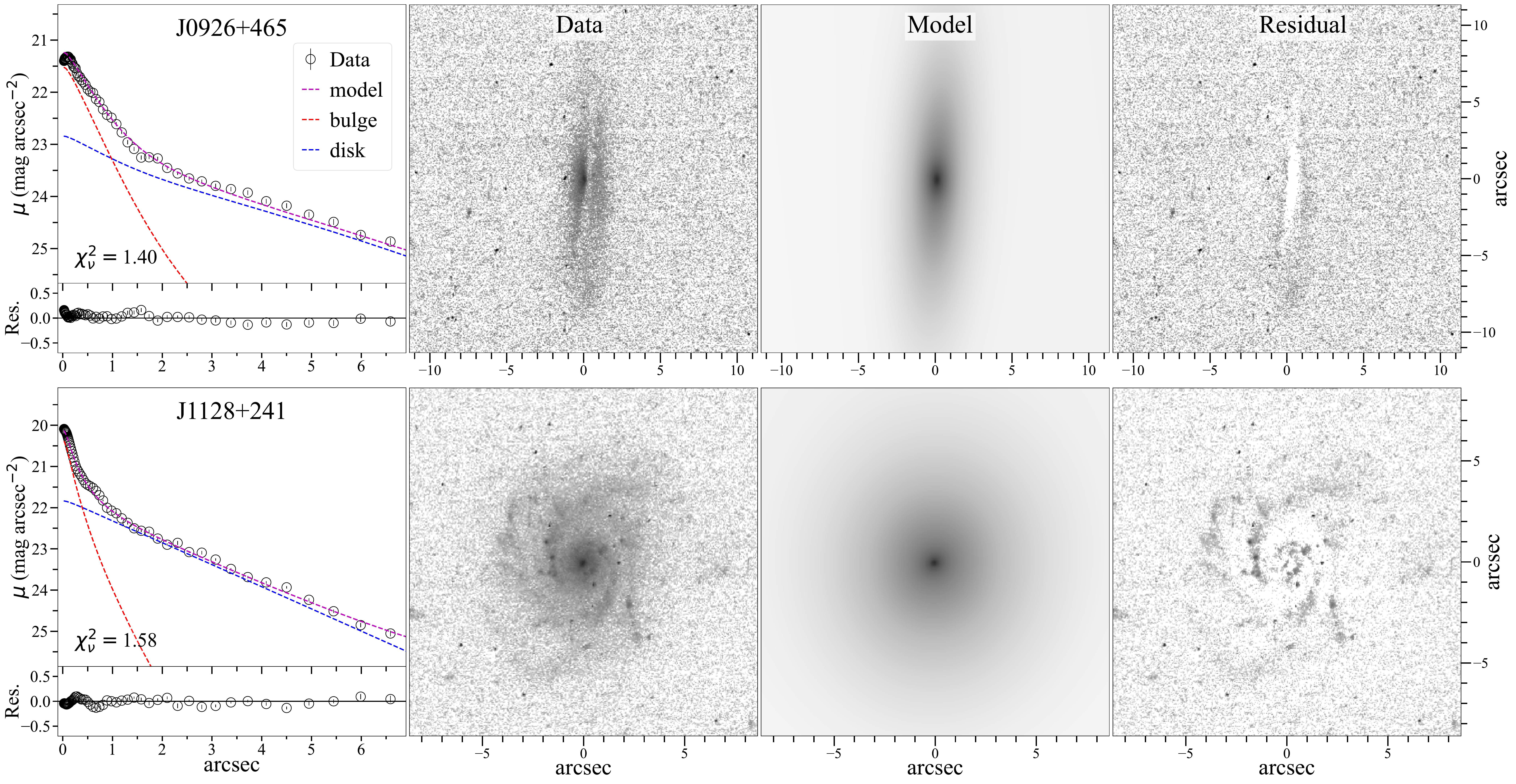}
\caption{Examples of two-dimensional bulge-disk decomposition using \GALFIT. The left column shows the surface brightness profile of the data (open circles with error bars), \sersic \ bulge component (red), exponential disk component (blue), and total (bulge + disk) model (purple). The $\chi^2_\nu$ from \GALFIT \ is shown in the lower-left corner. The lower panel gives the residuals between the data and the model. The right three columns show the original data, best-fit total model, and residuals, respectively. All images are displayed on a log stretch.}
\label{fig:galfit}
\end{figure*}

\begin{deluxetable*}{cccccccccccccccc}
\tablecaption{Measurement of Host Galaxy Properties of the High-confidence Sample \label{tab:measure}}
\tablewidth{0pt}
\tablehead{
\colhead{Name} & \colhead{$m_\mathrm{tot}$} & \colhead{$r_e$} & \colhead{$n_\mathrm{global}$} & \colhead{$\epsilon$} & \colhead{$C$} & \colhead{$m_{\mathrm{bulge}}$} & \colhead{$r_{\mathrm{bulge}}$} & \colhead{$\langle\mu_e\rangle_{\rm bulge}$} & \colhead{$n_{\mathrm{bulge}}$} &\colhead{PA$_{\mathrm{bulge}}$}&\colhead{$m_{\mathrm{disk}}$}&\colhead{PA$_\mathrm{disk}$}&\colhead{$B/T$}&\colhead{$\log\, M_*$} & \colhead{$g-r$}\\
\colhead{} & \colhead{(mag)} & \colhead{(\arcsec)} & \colhead{} & \colhead{} & \colhead{} & \colhead{(mag)} & \colhead{(\arcsec)} & \colhead{} & \colhead{} & \colhead{($^\circ$)}& \colhead{(mag)}& \colhead{($^\circ$)}& \colhead{}& \colhead{(${M_\odot}$)} & \colhead{(mag)}
}
\colnumbers
\startdata
J0209+075 & 19.07 &  2.76 & \nodata & 0.08 & 3.11 &   21.26 &    0.39 &   19.09 &    2.23 &     147 &   18.96 &     130 &    0.11 & $11.42\pm0.18$ & 0.86 \\
J0219+015 & 14.85 & 12.41 &    2.01 & 0.54 & 3.46 &   16.03 &    5.67 &   20.68 &    1.31 &     127 &   15.16 &     137 &    0.31 & $11.36\pm0.15$ & 0.84 \\
J0802+115 & 19.05 &  1.55 &    2.18 & 0.77 & 4.11 & \nodata & \nodata & \nodata & \nodata & \nodata & \nodata & \nodata & \nodata & $10.82\pm0.20$ & 0.68 \\
J0806+062 & 18.53 &  2.94 &    0.93 & 0.45 & 2.66 &   22.44 &    0.46 &   21.41 &    0.37 &     165 &   18.56 &     155 &    0.03 &  $9.81\pm0.52$ & 0.40 \\
J0832+184 & 17.17 &  3.12 &    3.50 & 0.42 & 4.53 &   18.06 &    1.01 &   18.70 &    1.81 &      18 &   17.62 &     164 &    0.40 & $11.27\pm0.15$ & 0.85 \\
J0847+124 & 18.24 &  5.85 &    1.28 & 0.66 & 3.00 &   21.29 &    1.14 &   21.92 &    0.53 &      56 &   18.22 &     115 &    0.06 & $11.33\pm0.15$ & 0.85 \\
J0855+420 & 19.70 &  2.04 &    2.06 & 0.50 & 3.17 &   20.75 &    1.05 &   21.12 &    1.45 &       2 &   20.26 &     173 &    0.39 & $11.07\pm0.20$ & 0.76 \\
J0914+413 & 16.77 & 12.00 &    3.89 & 0.26 & 3.24 &   18.06 &    2.06 &   20.16 &    2.45 &       5 &   17.02 &      27 &    0.28 & $11.85\pm0.15$ & 0.86 \\
J0926+465 & 19.66 &  7.30 &    2.66 & 0.74 & 3.32 &   22.55 &    0.61 &   21.69 &    0.48 &      68 &   19.73 &      68 &    0.07 & $10.92\pm0.25$ & 0.72 \\
J0956+162 & 20.53 &  2.21 &    0.87 & 0.64 & 3.08 &   22.66 &    0.46 &   20.43 &    0.62 &     167 &   20.50 &      23 &    0.12 & $10.78\pm0.25$ & 0.60 \\
J0958+561 & 19.13 &  4.76 &    5.53 & 0.47 & 4.35 &   19.90 &    0.83 &   19.45 &    2.85 &      91 &   19.59 &      95 &    0.43 & $11.17\pm0.17$ & 0.76 \\
J1128+241 & 18.45 &  4.16 &    2.23 & 0.19 & 2.93 &   21.07 &    0.54 &   20.35 &    1.72 &       3 &   18.40 &      50 &    0.08 & $10.30\pm0.23$ & 0.46 \\
J1328+571 & 16.74 &  9.42 &    1.63 & 0.41 & 3.27 & \nodata & \nodata & \nodata & \nodata & \nodata &   17.09 &      70 &    0.00 &  $8.89\pm1.34$ & 0.20 \\
J1633+084 & 19.68 &  3.53 &    1.66 & 0.49 & 2.99 &   21.86 &    0.57 &   20.42 &    0.37 &      45 &   19.64 &      47 &    0.11 & $11.35\pm0.19$ & 0.85 \\
J1646+383 & 17.75 &  2.28 &    2.76 & 0.35 & 3.76 &   19.63 &    0.50 &   18.80 &    0.16 &      26 &   18.06 &      32 &    0.19 & $10.91\pm0.15$ & 0.72 \\
J1656+640 & 18.33 &  4.68 &    2.13 & 0.56 & 3.43 &   19.91 &    1.43 &   20.83 &    1.08 &      54 &   18.50 &      56 &    0.21 & $11.46\pm0.15$ & 0.70 \\
J1721+262 & 18.63 &  3.44 & \nodata & 0.15 & 4.47 &   19.45 &    1.52 &   20.71 &    6.24 &     176 &   18.29 &      16 &    0.34 & $10.99\pm0.15$ & 0.78 \\
J2141+082 & 20.45 &  2.22 &    3.19 & 0.67 & 3.56 &   22.61 &    0.18 &   18.45 &    1.41 &      25 &   20.60 &      16 &    0.14 & $11.13\pm0.21$ & 0.53 \\
\enddata
\tablecomments{Col. (1): Object name. Col. (2): Petrosian magnitude after correction for Galactic extinction. Col. (3): Effective radius from fitting a single-\sersic \ model. Objects J0209+075 and J1721+262 are not well fit by a single-\sersic \ model, and for them we adopt the half-light radii from non-parametric measurements. Col. (4): Global \sersic \ index from the single-\sersic \ model fit. Col. (5): Ellipticity from non-parametric measurements, defined by $\epsilon = 1-b/a$. Col. (6): Concentration parameter, defined by $C=5\log\,({r_{80}/r_{20}})$. Cols. (7)--(13): Magnitude, effective radius, effective surface brightness ($R$\,mag\,arcsec$^{-2}$), \sersic \ index, and position angle (east of north) of the bulge and disk components derived from a two-component \sersic \ model. Col. (14): Bulge-to-total luminosity ratio. Col. (15): Stellar mass. Col. (16): Rest-frame $g-r$ color of the entire galaxy. Object J0802+115 is an irregular galaxy without meaningful bulge-disk decomposition, while object J1328+571 is a dwarf (Magellanic spiral) galaxy without a bulge ($B/T=0$).}
\end{deluxetable*}

\subsection{Properties of the Host Galaxy}
\label{sec:nonpara}
We first estimate and subtract the sky background of the HST images using a sigma-clipping method that iteratively clips the images beyond $3\,\sigma$, where $\sigma$ is the standard deviation of the background pixel distribution. We use \Phot \ to create a segmentation map, setting the segmentation threshold to $2\,\sigma$ above the background noise. The segmentation map allows us to improve the background measurement by calculating the pixel distribution in the sky regions. We create a mask image to remove unrelated sources according to the segmentation map.

With the sky-subtracted image and corresponding segmentation map and mask image, we use the {\tt Python} package \texttt{statmorph} \citep{Rodriguez-Gomez_2019_MNRAS_483_4140} to compute several non-parametric quantities of interest (Table~\ref{tab:measure}), including the total magnitude ($m_{\rm tot}$) within the \cite{Petrosian_1976_ApJL_210_L53} radius, ellipticity $\epsilon \equiv 1-b/a$, with $a$ the semi-major and $b$ the semi-minor axes calculated from the second-order moments\footnote{Equations~24--25 from the {\tt SExtractor} user manual at \url{https://readthedocs.org/projects/sextractor/downloads/}}, and the concentration parameter $C\equiv5 \log\,{r_{80}/r_{20}}$ \citep{Conselice_2003_ApJS_147_1}, where $r_{80}$ and $r_{20}$ are the circular aperture radii enclosing 80\% and 20\% of the total flux, respectively.

We use \GALFIT\ to model the two-dimensional light distribution of the galaxy. As in \citet{Zhuang_Ho2022}, we construct a point-spread function using \texttt{reproject}\footnote{https://reproject.readthedocs.io/en/stable/index.html} to stack unsaturated bright stars within the same image of each object. To reproject and stack the stars, we adopt the flux-conserving scheme \texttt{spherical polygon intersection}, which treats pixels as four-sided spherical polygons and computes the exact overlap of pixels on the sky. We first fit the galaxy using a single-\sersic\ component to obtain a global model, yielding the effective radius $r_e$ and \sersic\ index $n_{\rm global}$. We then fit two components to decompose the bulge and disk\footnote{Although Zoo Gems also used \GALFIT\ to analyze the images, no details were given by \cite{Keel_2022_AJ_163_150}.}, using a \sersic\ component with a free index $n_{\rm bulge}$ to model the bulge and an exponential profile to model the disk. Obvious dust lanes were masked manually to mitigate their effects on the fits. The model residual images show that in general the two-component bulge-disk decompositions are quite successful. We find no evidence for any additional component that might arise from a prominent nucleus. Figure~\ref{fig:galfit} shows a couple of examples of the decompositions, and the fits for the full set of galaxies can be found in Appendix~\ref{sec:galfit_img}. Table~\ref{tab:measure} summarizes the fitting results, which include quantitative parameters for the bulge and disk. Our fits do not take into account detailed features such as spiral arms. According to \cite{Gao_2017_ApJ_845_114}, the systematic uncertainties introduced by this over-simplification are 0.14\,mag for $m_{\rm bulge}$, 10\% for $r_{\rm bulge}$, and 14\% for $n_{\rm bulge}$.

We estimate the stellar mass of the host galaxies using SDSS multiband photometry and the method of \cite{Taylor_2011_MNRAS_418_1587}, which is based on the rest-frame $i$-band absolute magnitude and $g-i$ color of $z < 0.65$ (median $z = 0.2$) galaxies. The $1 \,\sigma $ uncertainty in $\log\, M_*$ is $\sim 0.1$ dex. We use SDSS magnitudes in the $g$ and $i$ bands and apply K-correction \citep{Chilingarian_2010_MNRAS_405_1409, Chilingarian_2012_MNRAS_419_1727} and Galactic extinction correction to convert to rest-frame magnitudes. The final error budget of the stellar mass has contributions from the redshift, K-correction, and the stellar mass estimation method. Seven objects overlap with the second release of the GALEX-SDSS-WISE Legacy Catalog \citep[GSWLC-2;][]{Salim_2016_ApJS_227_2, Salim_2018_ApJ_859_11}, whose stellar masses are derived from ultraviolet-to-infrared SED fitting. Comparing the stellar masses estimated in this paper and those from GSWLC-2, we find good agreement between the two methods, with maximum difference $\Delta \log\, M_*<0.3$ dex.

\begin{figure*}[]
\plotone{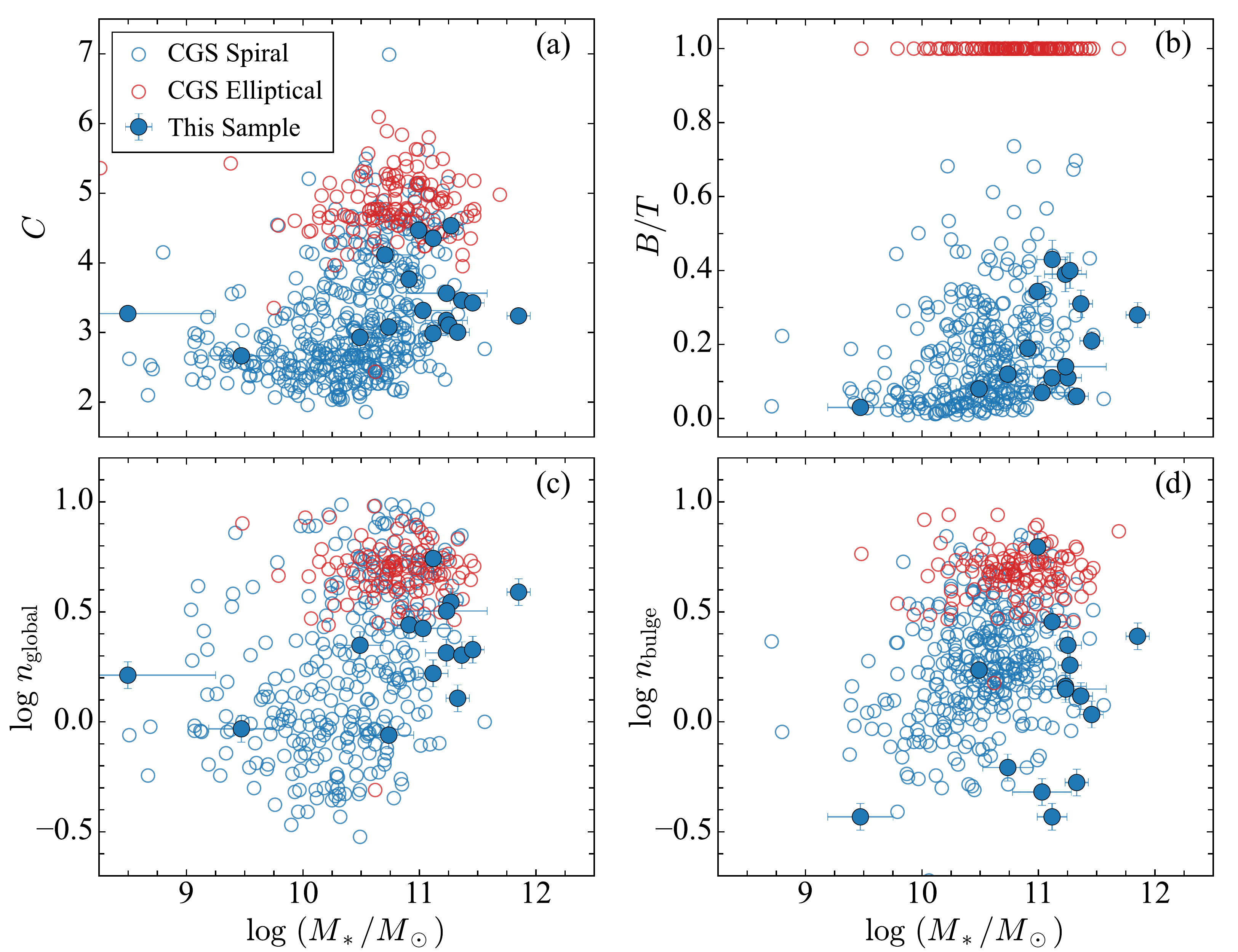}
\caption{Distribution of (a) concentration parameter $C$, (b) bulge-to-total light ratio $B/T$, (c) global \sersic\ index $n_\mathrm{global}$, and (d) bulge \sersic\ index $n_\mathrm{bulge}$ for our sample of disk radio galaxies, in comparison with the control sample of normal galaxies from CGS \citep[blue symbols, spirals; red symbols, ellipticals;][]{Li_2011_ApJS, Gao_2019_ApJS}. The irregular galaxy J0802+115 and the dwarf Magellanic spiral galaxy J1328+571, which do not have a bulge, are not shown in panels (b) and (d). All quantities pertain to the $R$ band.}
\label{fig:dist}
\end{figure*}

We compare the radio galaxies with normal galaxies in CGS \citep{Ho_2011_ApJS_197_21}, a statistically complete sample of 605 bright ($B_T<12.9\,$mag), southern ($\delta < 0\degr$) galaxies imaged in the optical with the facilities at Las Campanas Observatory. The images have a median seeing of $\sim 1\arcsec$, field-of-view of $8\farcm9\times 8\farcm9$, and median limiting surface brightness $\sim 27.5$, 26.9, 26.4, and 25.3\,mag\,arcsec$^{-2}$ in the $B$, $V$, $R$, and $I$ bands, respectively. The flux fraction radii (thus concentration parameter) are provided in \cite{Li_2011_ApJS}, and \cite{Gao_2019_ApJS} give bulge-to-total light ratios ($B/T$) and bulge \sersic \ indices. We measure the global \sersic \ index in the $B$ band with \GALFIT, as in Section~\ref{sec:nonpara}. The stellar masses of the CGS galaxies are obtained from modeling with a delayed star formation history using {\tt CIGALE} \citep{Boquien+2019A&A} their complete optical and near-IR photometry (M.-Y. Zhuang et al., in preparation), when available, and otherwise from $B-V$ color and $K$ or $I$ (if $K$ is not available) magnitude using conversions from \cite{Bell_2003_ApJS}. Figure~\ref{fig:dist} shows parameter distributions of the radio disk galaxies compared to the control sample of CGS galaxies. Although our galaxies were not observed in the same bands and have slightly higher redshift ($z\approx0.2$) than the CGS nearby galaxy sample, the difference in structural parameters is expected to be mild \citep{Haubler_2013_MNRAS, Haubler_2022_arXiv, Conselice_2014_ARAA}.

\begin{figure*}[]
\plottwo{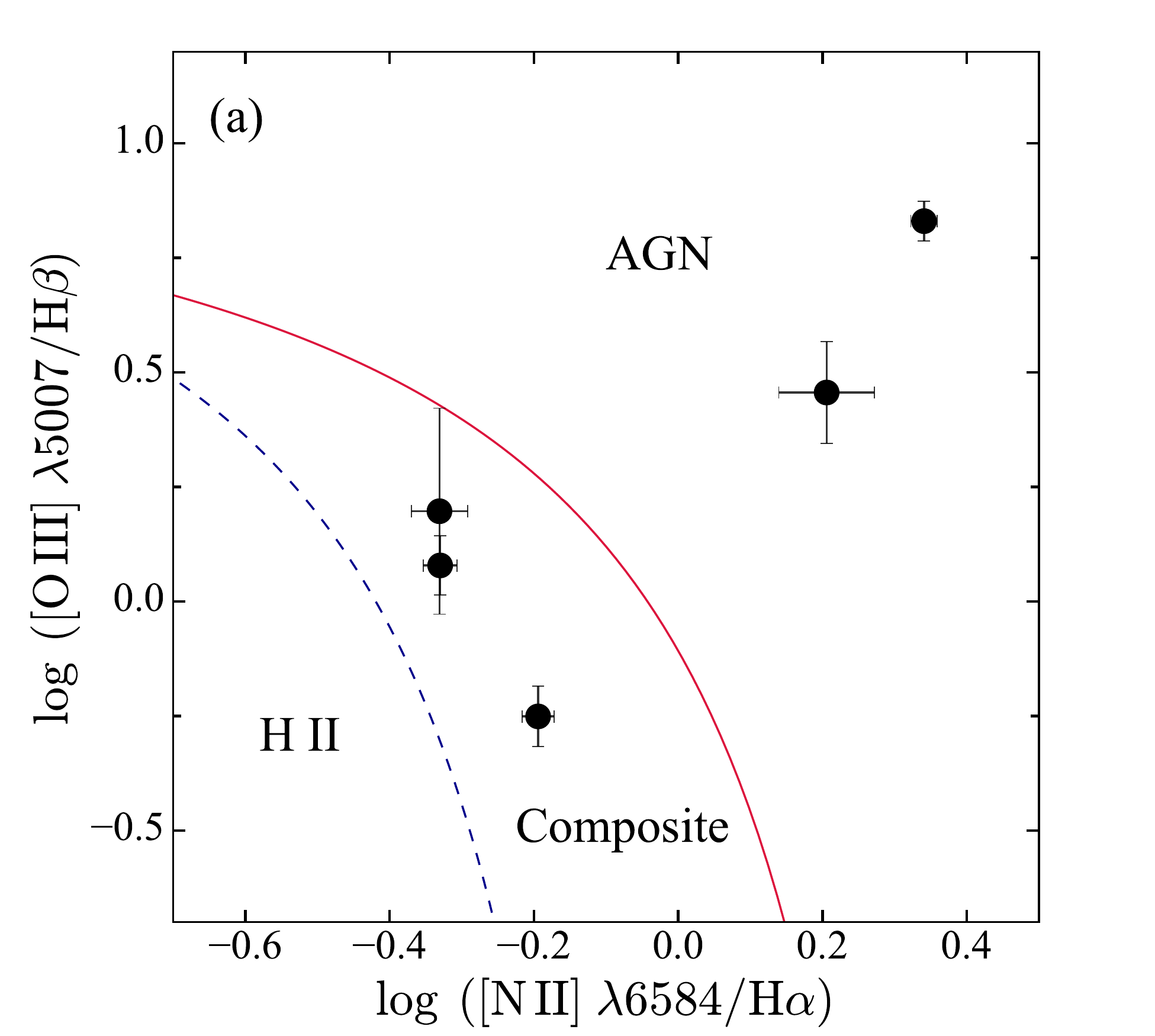}{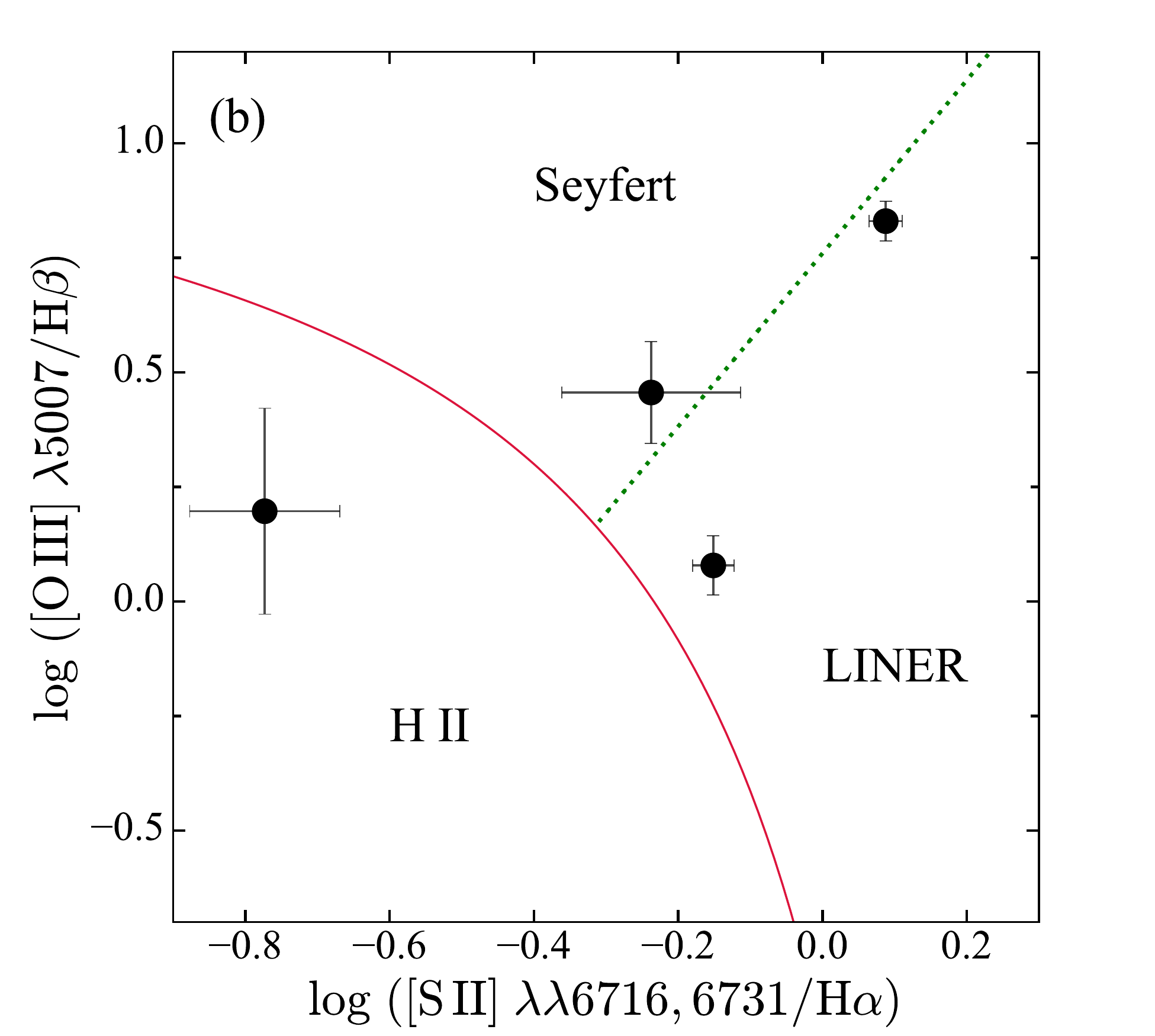}
\caption{The \OIII/H$\beta$ versus (a) \NII/H$\alpha$ and (b) \SII/H$\alpha$ diagnostic diagrams for objects with signal-to-noise ratio larger than 2 in their optical emission-line fluxes. The extreme starburst boundary (red solid line; \citealt{Kewley_2001_ApJ}), the pure star formation line (blue dashed line in panel a; \citealt{Kauffmann_2003_MNRAS_346_1055}), and the separation between Seyferts and LINERs (green dotted line in panel b; \citealt{Kewley_2006_MNRAS_372_961}) are used to classify galaxies into star-forming (H$\,$II) galaxies, Seyferts, LINERs, and composite (star-forming and AGN) systems.}
\label{fig:BPT}
\end{figure*}

Six galaxies in the high-confidence sample have SDSS spectra, whose spectroscopic fiber covers the central 3\arcsec\ or $\sim 2-10\,$kpc of the galaxies. We obtain emission-line fluxes from the SDSS \texttt{galSpecLine} catalog \citep{Brinchmann_2004_MNRAS_351_1151}. For objects with signal-to-noise ratios larger than 2 for H$\beta$, \OIII~$\lambda$5007, H$\alpha$, and \NII~$\lambda 6584$ or \SII~$\lambda\lambda6716, 6731$, we show their \cite{Baldwin_1981_PASP_93_5} diagnostic diagrams in Figure~\ref{fig:BPT}. Five objects host AGNs or composite (AGN plus star-forming) nuclei according to the \OIII/H$\beta$--\NII/H$\alpha$ diagram, while in the \OIII/H$\beta$--\SII/H$\alpha$ diagram one object is a Seyfert and two are low-ionization nuclear emission-line regions (LINERs; \citealt{Heckman_1980_A&A}). The two LINERs (J0914+413 and J1721+262) both have a compact radio core in their nucleus, a common feature in AGNs with low accretion rates \citep{Ho_2002_ApJ_564_120, Ho2008}.

\begin{figure}[b]
\includegraphics[width=0.47\textwidth]{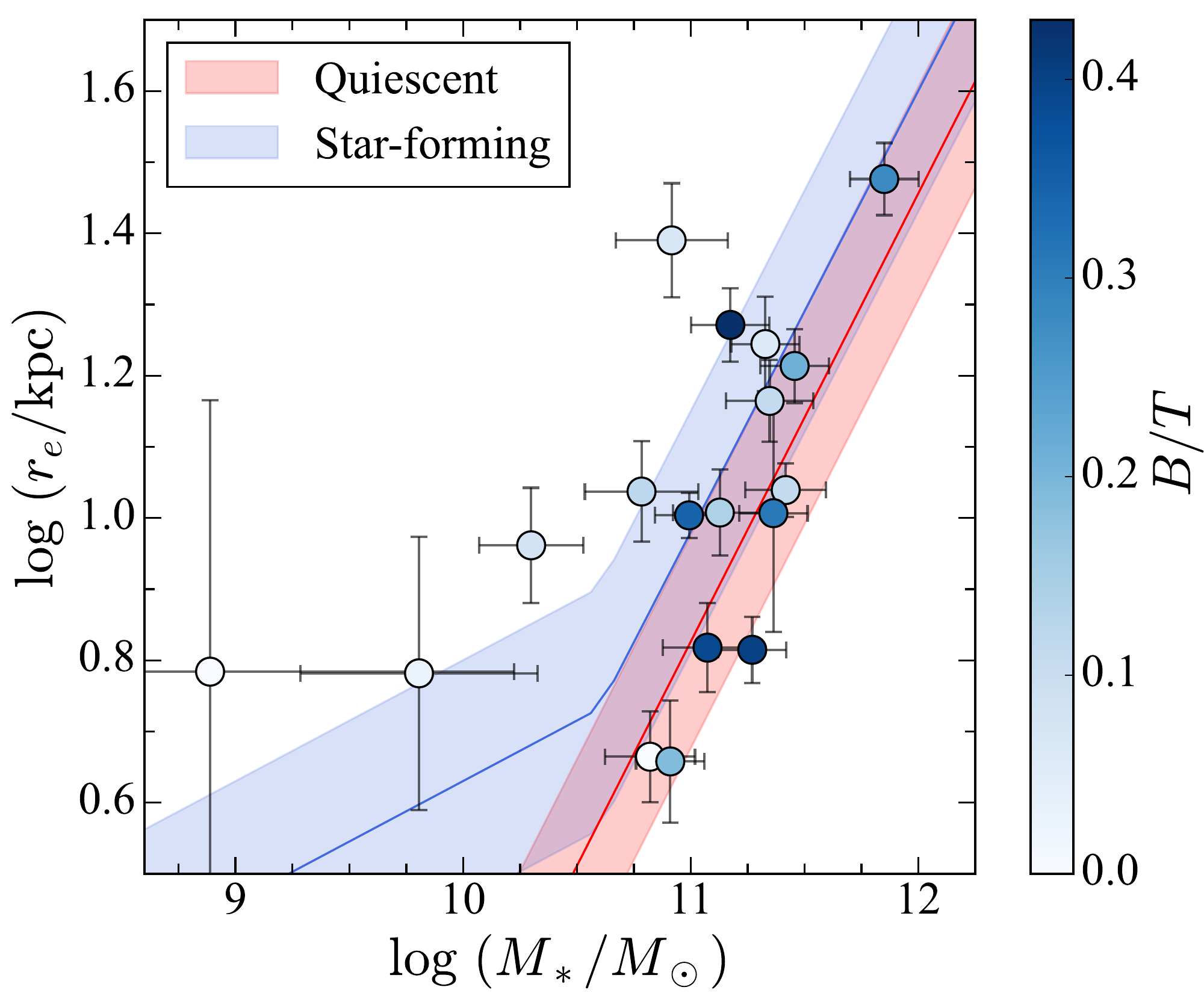}
\caption{Effective radius ($r_e$) versus stellar mass ($M_*$) for the host galaxies of the double-lobed radio sources in our sample, with colors indicating their bulge-to-total light ratios ($B/T$). Blue and red solid lines represent the mass-size relations of star-forming and quiescent galaxies at $z \approx 0.2-0.4$ \citep{Kawinwanichakij_2021_ApJ}, with shaded areas indicating the $1\,\sigma$ intrinsic scatter of $\log\, r_e$.}
\label{fig:mass_size}
\end{figure}

\begin{deluxetable*}{lcccrrccrcc}[t]
\tablecaption{Double-lobed Radio Sources Hosted by Disk Galaxies from the Literature \label{tab:literature}}
\tablewidth{0pt}
\tablehead{
\colhead{Name} & \colhead{R. A. (J2000)} & \colhead{Decl. (J2000)} & \colhead{$z$} & \colhead{$D$} & \colhead{$L_{\mathrm{1.4}}$} & \colhead{FR Type} & \colhead{$M_R$} & \colhead{$\log\,M_\mathrm{*}$} & \colhead{$\log\,M_\mathrm{BH}$} & \colhead{Ref.} \\ \colhead{} & \colhead{($^{\mathrm{h}}\;{}^{\mathrm{m}}\;{}^{\mathrm{s}}$)} & \colhead{($^\circ\;{}^\prime\;{}^{\prime\prime}$)} & \colhead{} & \colhead{(kpc)} & \colhead{($\mathrm{W\, Hz^{-1}}$)} & \colhead{} &\colhead{(mag)} & \colhead{($M_\odot$)} & \colhead{($M_\odot$)} & \colhead{} \\
\colhead{(1)} &
\colhead{(2)} &
\colhead{(3)} &
\colhead{(4)} &
\colhead{(5)} &
\colhead{(6)} &
\colhead{(7)} &
\colhead{(8)} &
\colhead{(9)} &
\colhead{(10)} &
\colhead{(11)}
}
\startdata
J0315$-$1906 (0313$-$192) & 03 15 52.1 & $-$19 06 44 & 0.067 &  200 & 1.0 $\times10^{24}$  &  I & $-21.28$ & 10.41 & 7.31 &  1 \\
J0354$-$1340              & 03 54 32.8 & +13 40 07   & 0.076 &  240 & $2.06\times 10^{23}$ & II & $-21.91$ & 10.66 & 7.72 & 2, 3\\
J0836$+$0532              & 08 36 55.9 & +05 32 42   & 0.099 &  420 & $1.53 \times10^{24}$ & II & $-23.11$ & 11.36 & 8.85 &  4 \\
J1159$+$5820              & 11 59 05.8 & +58 20 36   & 0.054 &  392 & $2.26 \times10^{24}$ & II & $-23.32$ & 11.36 & 8.85 &  4 \\
J1352$+$3126              & 13 52 17.8 & +31 26 46   & 0.045 &  335 & $2.26 \times10^{25}$ & II & $-22.80$ & 11.32 & 8.79 &  4 \\
J1409$-$0302 (Speca)      & 14 09 48.8 & $-$03 02 32 & 0.138 & 1000 & 7.0 $\times10^{24}$  & II & $-23.16$ & 11.57 & 9.19 &  5 \\
J1649$+$2635              & 16 49 23.9 & $+$26 35 03 & 0.055 &   86 & $1.07 \times10^{24}$ & II & $-22.85$ & 11.42 & 8.94 &  4, 6 \\
J2345$-$0449              & 23 45 32.7 & $-$04 49 25 & 0.076 & 1600 & $2.5 \times10^{24}$  & II & $-22.86$ & 11.04 & 8.33 &  7 \\
MCG+07$-$47$-$10          & 23 18 32.7 & +43 14 49   & 0.012 &  207 & $1.12 \times 10^{22}$& II & $-20.67$ & 10.16 & 6.92 &  8, 9 \\
\enddata
\tablecomments{ Col. (1): Object name. Cols. (2)--(3): Coordinates. Col. (4): Redshift. Col. (5): Projected proper distance between the two radio hotspots; for J0354$-$1340, what is shown is the de-projected linear size. Col. (6): Radio luminosity at 1.4\,GHz. Col. (7): FR type. They are predominantly classified as FR~II but have much lower radio power than the lower limit of FR~II sources hosted by elliptical galaxies \citep[$3\times 10^{25}\;\mathrm{W\;Hz^{-1}}$;][]{Tadhunter_2016_AAPR_24_10}. Col. (8): Host galaxy $R$-band absolute magnitude. For objects J0836+0532, J1159+5820, J1352+3126, J1409$-$0302, and J1649+2635, we derive the $R$-band magnitude from the SDSS $gri$ magnitudes according to \cite{Jester_2005_AJ_130_873}. For the others with more limited photometry, we derive their $R$-band magnitude assuming the color of Sab galaxies, following the conversions in \cite{Fukugita_1995_PASP_107_945} and \cite{Chilingarian_2017_ApJS}. Col. (9): Stellar mass of the host galaxy, computed as described in Section~4.2. Col. (10): Black hole mass, computed from $M_*$ as described in Section~5.2. Col. (11): References: (1) Ledlow et al. 1998; (2) Chen et al. 2020; (3) Vietri et al. 2022; (4) Singh et al. 2015; (5) Hota et al. 2011; (6) Mao et al. 2015; (7) Bagchi et al. 2014; (8) Mulcahy et al. 2016; (9) Bilicki et al. 2014.}
\end{deluxetable*}

\begin{figure}[b]
\includegraphics[width=0.48\textwidth]{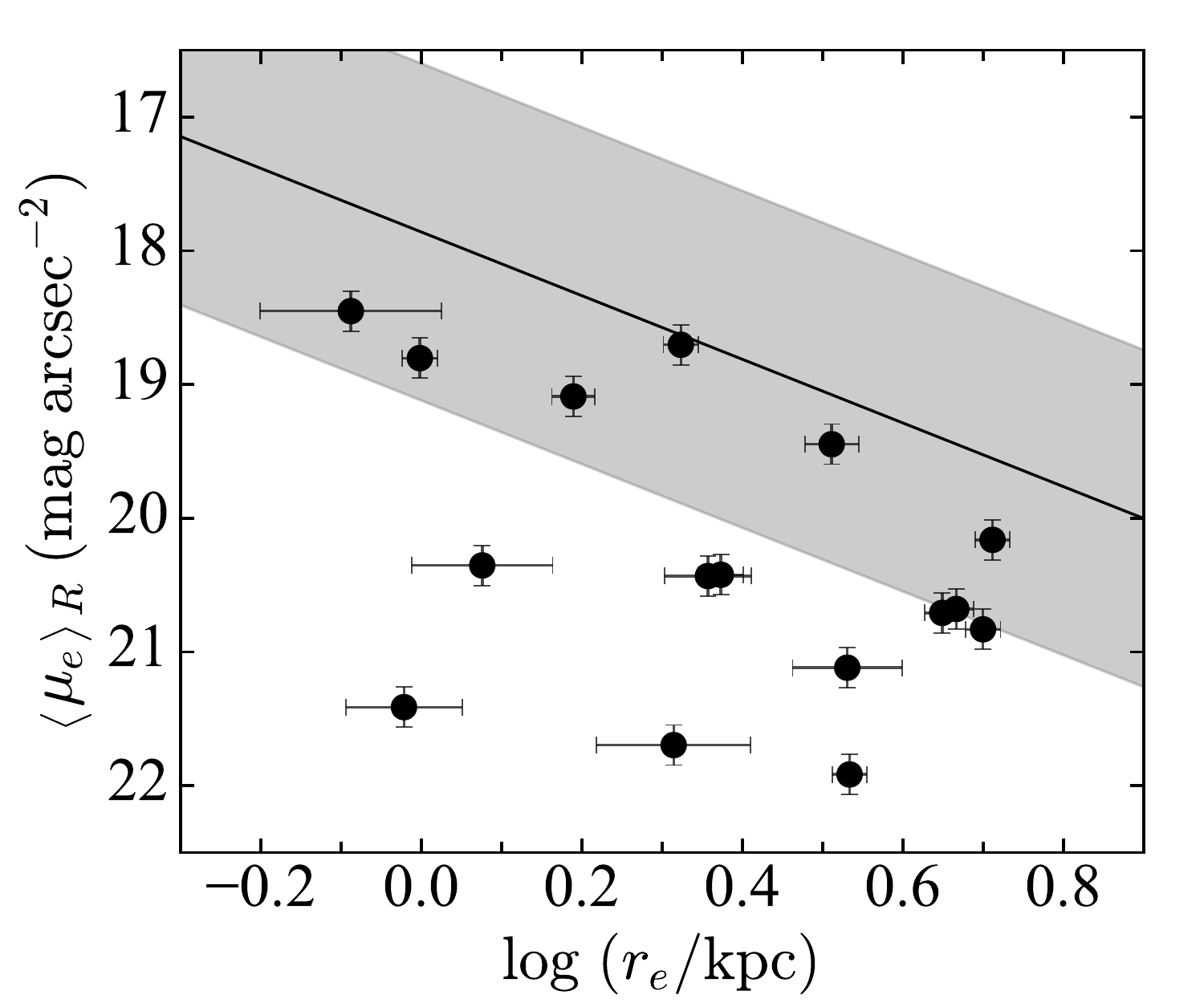}
\caption{Distribution of bulges of host galaxies in the Kormendy relation. The black solid line shows the Kormendy relation measured in the $R$ band from \cite{Gao_2020_ApJS}, with the shaded region showing the $3\,\sigma$ range. Classical bulges and ellipticals follow the Kormendy relation, while pseudo bulges are low-surface brightness outliers below the relation. The HST F475W magnitudes have been K-corrected to the $R$ band.}
\label{fig:Kormendy}
\end{figure}

\section{Statistical Properties}
\label{sec:results}

\subsection{Massive, Quiescent, Disk-dominated Hosts}

Radio galaxies mostly consist of giant ellipticals with stellar mass exceeding $10^{11}\,M_\odot$ \citep{Best_2005_MNRAS_362_25}. The hosts of our sample of double-lobed radio galaxies are also predominantly massive systems (Figure~\ref{fig:dist}), having a median stellar mass $M_* = 1.3 \times 10^{11}\,M_\odot$. As with other massive galaxies in the nearby Universe, they have optical colors ($\langle g-r \rangle = 0.69 \pm 0.18\,$mag, corrected for redshift and Galactic extinction) that place them on the red sequence \citep{Blanton2009}. There are, however, two noteworthy outliers that have blue colors and unexpectedly low stellar masses: J0806+062 with $M_* = 6.5\times 10^9\,M_\odot$ and $g-r = 0.4\,$mag, and J1328+571 with $M_* = 7.8\times10^8\,M_\odot$ and $g-r = 0.13\,$mag. The extreme late-type, barred spiral morphology of J1328+571 bears a striking resemblance to the Large Magellanic Cloud, which is only $\sim 3$ times more massive ($M_* = 2.7\times 10^9\,M_\odot$; \citealt{vanderMarel2002}). Nothing particularly unusual stands out in terms of the effective radii of the sample: most of the members obey the stellar mass-size relation of massive, quiescent galaxies at $z \approx 0.2-0.4$ (Figure~\ref{fig:mass_size}; \citealt{Kawinwanichakij_2021_ApJ}).

However, completely counterintuitive to expectation, the galaxies in our sample have stellar morphologies and internal substructures of unmistakably disk-dominated, in many instances extremely late-type galaxies. Even a cursory inspection of the high-quality, optical HST images in Figure~\ref{fig:figset} reveals that the hosts clearly have disks, as evidenced by spiral structure when viewed at low inclinations and large-scale dust lanes when seen edge-on. Six galaxies (J0209+075, J0806+062, J1128+241, J1328+571, J1646+383, J1656+640) have spiral features, and seven (J0219+015, J0847+124, J0855+420, J0914+413, J0926+465, J0956+162, J1633+084) have prominent dust lanes with dimensions comparable to that of the entire galaxy, which are common in spiral galaxies and are regarded as signatures of edge-on spiral disks \citep{Holwerda_2019_AJ_158_103}. In sum, we conclude that $\sim 70\%$ of the sample has spiral arms. The late-type nature of the hosts is further supported by more quantitative metrics. For instance, the sample galaxies have concentrations (median $C = 3.3 \pm 0.5$) consistent with late-type galaxies (\citealt{Conselice_2014_ARAA}; $C = 3.1 \pm 0.4$), which is further reinforced by their high global ellipticities (median $\epsilon = 0.48 \pm 0.19$) and low \sersic\ indices (median $n_{\rm global} = 2.2 \pm 1.2$). Such low values of optical concentration and global \sersic\ index deviate strongly from the majority of $M_* \approx 10^{11}\,M_\odot$ galaxies (Figures~\ref{fig:dist}a and \ref{fig:dist}c), and elliptical galaxies rarely are flatter than $\epsilon = 0.7$ \citep{Sandage1970}.

An equally remarkable testimony to the exceptionally late-type nature of the host galaxies comes from the bulge-to-disk decomposition analysis described in Section~4.2 (Figure~\ref{fig:galfit}; Appendix~C), which yields a sample median $B/T = 0.13$. Seven sources have $B/T \lesssim 0.1$ and can be deemed essentially bulgeless. Formally speaking, J1328+571 is truly bulgeless. As a consequence of their different evolutionary histories \citep{Kormendy_2004_ARAA, Kormendy_2013_ARAA_51_511}, galactic bulges fall into two types---classical and pseudo---that broadly can be distinguished by their internal kinematics, structure, and radial light profile. Many authors (e.g., \citealt{Fisher_2008_AJ_136_773, Fisher_2016}) classify bulges according to the value of their \sersic\ index: classical and pseudo bulges are defined by $n > 2$ and $n \leq 2$, respectively. By this criterion, the vast majority (12/16 or 75\%) of our sample galaxies with successful bulge-to-disk decomposition qualify as having a pseudo bulge, not an unanticipated result in view of the low $B/T$ values that dominate the sample and the tendency for pseudo bulges to have $B/T \lesssim 0.35$ \citep{Kormendy_2013_ARAA_51_511, Kormendy2016}. Others argue, however, that the \sersic\ index can yield misleading bulge classifications (e.g., \citealt{Gadotti2009, Gao2018}), advocating, instead, that bulge classification should place greater reliance on the \cite{Kormendy_1977_ApJ} relation between surface brightness and effective radius, to which classical bulges and ellipticals adhere but pseudo bulges do not \citep{Neumann2017, Gao_2020_ApJS, Gao2022, Sachdeva2020}. Using the CGS $R$-band Kormendy relation of ellipticals and classical bulges as a reference, Figure~\ref{fig:Kormendy} illustrates that the bulges of our sample of disky radio galaxies systematically deviate to lower surface brightness at fixed effective radii, consistent with the expected behavior of pseudo bulges \citep{Gao_2020_ApJS, Gao2022}. To enable this comparison, we K-corrected the HST F475W magnitudes to the $R$ band, assuming a bulge SED from \cite{Kinney_1996_ApJ} and using {\tt Astrolib PySynphot} \citep{2013ascl.soft03023S}.

It is quite unusual for spiral galaxies to have stellar masses as large as those in our sample, for the characteristic stellar mass of spiral galaxies is $\sim 3\times 10^{10}\,M_\odot$, above which the galaxy number density significantly drops \citep{Kelvin_2014_MNRAS_444_1647, Ogle_2016_ApJ_817_109}. Spiral galaxies with $M_* > 10^{11}\,M_\odot$ are extremely uncommon and have been systematically studied only recently \citep{Ogle_2016_ApJ_817_109, Ogle_2019_ApJS_243_14}. Rarer still are spiral galaxies that host powerful radio AGNs. The large fraction of massive objects in our sample suggests that they are not a random subset of normal galaxies. The nearest neighbors two-sample test \citep{Rizzo_2019}, as implemented in the \texttt{CRAN} package \texttt{yaImpute} \citep{Crookston_2008}, yields $p$-values $<$ 0.02 when we compare their distributions with CGS spiral galaxies in the four diagrams of Figure~\ref{fig:dist}. Prior to this study, nine secure cases of double-lobed radio sources hosted by disk had been reported (Table~5), among which only one (J0315$-$1906; \citealt{Keel2006}) had the benefit of high-resolution imaging from HST. Intriguingly, six of the nine cases also have stellar masses $M_* \gtrsim 10^{11} \,M_\odot$. Together with the statistics from our new sample, these trends suggest that the formation of large-scale radio lobes may be linked to the unusually high stellar mass of the hosts. We will revisit this topic in Section~6.

\begin{figure}[t]
\includegraphics[width=0.445\textwidth]{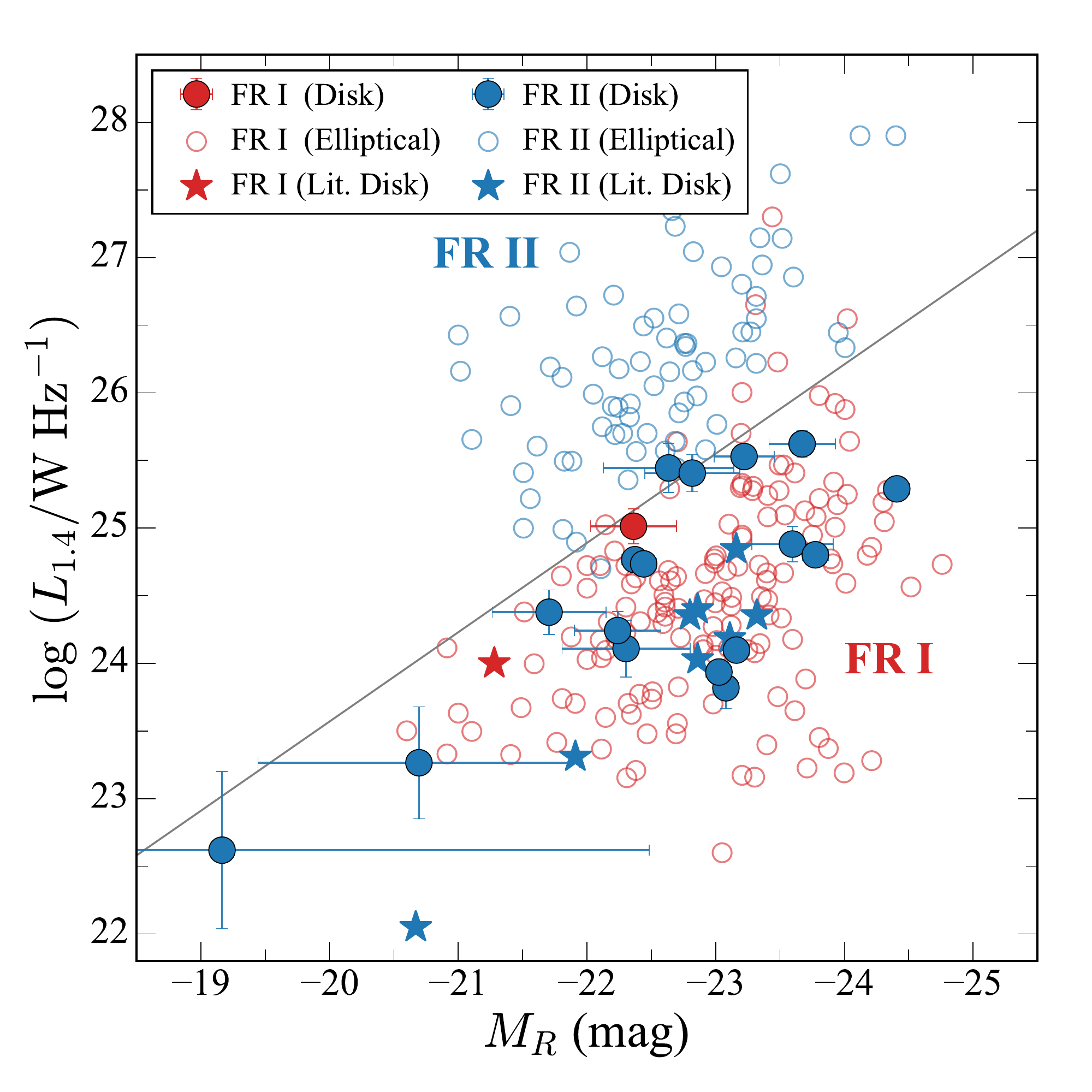}
\caption{Radio--optical luminosity relation showing the FR~I--FR~II dichotomy. The solid line is the Owen-Ledlow relation that divides FR~I (red) and FR~II (blue) sources. Large filled circles are objects from our sample, stars are objects from the literature (Table~5), and small open circles are elliptical galaxies from \cite{Owen_1994}.}
\label{fig:FR II}
\end{figure}

\begin{figure}[t]
\includegraphics[width=0.50\textwidth]{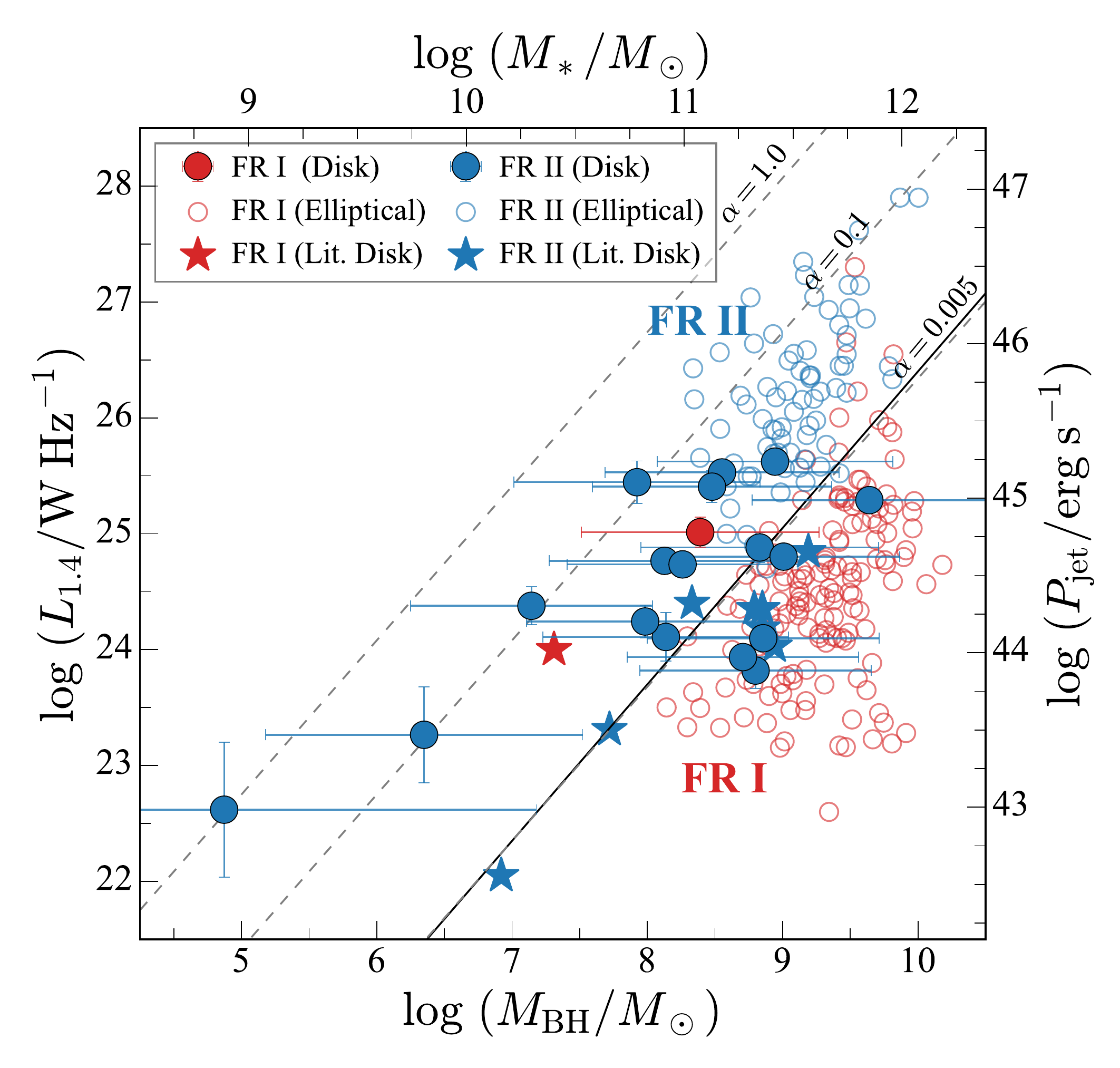}
\caption{Radio luminosity--BH mass relation showing the FR~I--FR~II dichotomy. The top $x$-axis gives the total stellar mass. The right $y$-axis converts radio luminosity to jet power following Equation~10. The solid line is the BH mass version of the Owen-Ledlow relation, where we converted the $R$-band absolute magnitude to BH mass as described in Section~5.2. Dashed lines show constant critical ratios of jet power over Eddington luminosity, $\alpha \equiv P_\mathrm{jet} / L_\mathrm{Edd} = 1$, 0.1, and 0.005. Large filled circles are objects from our sample, stars are objects from the literature (Table~5), and small open circles are elliptical galaxies from \cite{Owen_1994}.}
\label{fig:FR II_BH}
\end{figure}

\subsection{FR~I--FR~II Dichotomy}

FR~II radio sources are on average more powerful than FR~I radio sources \citep{Fanaroff_1974_MNRAS_167_31P}, with the division occuring at a power of $\sim 10^{26}\, \mathrm{W \, Hz}^{-1}$ at 178\,MHz, or, equivalently at 1.4\,GHz, $L_{1.4}\approx 3\times 10^{25}\,\mathrm{W\,Hz^{-1}}$ \citep{Tadhunter_2016_AAPR_24_10}. \citet[][see also \citealt{Ledlow_1996_AJ_112_9}]{Owen_1994} suggested that the division line correlates with the absolute magnitude of the host galaxy,

\begin{equation}
\log\, \left(\frac{L_\mathrm{1.4}}{{\rm W\,Hz^{-1}}} \right) = -0.66 M_{R}+10.37,
\label{eq:owen}
\end{equation}

\noindent
where $M_R$ is the Cousins $R$-band absolute magnitude. Figure~\ref{fig:FR II} shows where our objects lie with respect to the FR~I--FR~II dichotomy, as defined by the elliptical radio galaxy sample of \cite{Owen_1994}. We obtain the $R$-band magnitude of our objects from their SDSS magnitudes, adopting conversions from \cite{Jester_2005_AJ_130_873} with K-correction applied. To maximize the sample, we also add the previously known disk galaxies hosting double-lobed radio sources (Table~\ref{tab:literature}). 

Even though all but two of the combined sample of 27 objects have FR~II morphologies, nearly all lie below the Owen-Ledlow relation because they tend to have much lower radio power than traditional FR~II sources hosted by elliptical galaxies of the same optical luminosity. Violating the well-established FR~I--FR~II dichotomy, disk radio sources have radio powers characteristic of FR~I systems but nevertheless display FR~II morphologies. Moreover, the fraction of FR~II sources among spiral hosts significantly exceeds that in elliptical radio galaxies, for which the proportion of FR~I and FR~II types is roughly equal \citep{Capetti_2017_AA_598, Capetti_2017_AA_601}. \cite{Singh_2015_MNRAS_454_1556} suggest that the radio lobes hosted by spiral galaxies may be in a late phase of evolution, caught during a period when the radio emission has subsided. However, if radio lobes expand slowly and fade rapidly, as theory predicts \citep{Luo_2010_ApJ_713_398}, the fading stage lasts but a fleeting moment compared to the lifetime of the jet, rendering the fading scenario implausible to explain the low powers of all disk radio sources discovered to date.

What physically underpins the Owen-Ledlow relation? In view of the fundamental connection between BH mass and bulge stellar mass \citep{Kormendy_2013_ARAA_51_511}, \cite{Ghisellini_2001_AAP_379_L1} proposed that the Owen-Ledlow relation ultimately reflects a connection between radio power and the mass of the central BH. To revisit this idea, we estimate the BH masses of our combined sample of spiral hosts (Tables~4 and 5) with the aid of the empirical scaling relation between BH mass and galaxy total stellar mass \citep{Greene_2020_ARAA_58_257}. Note that the slope, zero point, and intrinsic scatter of the $M_\mathrm{BH}-M_*$ relation vary considerably depending on galaxy morphological type. While the disk-dominated structure of the hosts tempts us to consider the relation for late-type galaxies, we are conflicted simultaneously by the large stellar masses of the hosts that more closely mimic early-type galaxies. For concreteness, we choose the relation calibrated using all galaxy types combined:

\begin{equation}
\log\, \left(\frac{M_\mathrm{BH}}{M_\odot} \right) = (7.43 \pm0.09) + (1.61 \pm 0.12)\log\, \left(\frac{M_*}{M_0} \right),
\label{equation8}
\end{equation}

\noindent
with $M_0 = 3 \times 10^{10} \, M_\odot$ and an intrinsic scatter of 0.81~dex. At a fiducial stellar mass of $M_* = 10^{11}\,M_\odot$, the scaling relation of early-type galaxies overpredicts $M_\mathrm{BH}$ by 0.36~dex, while the late-type galaxy calibration underpredicts $M_\mathrm{BH}$ by 0.69~dex. We are even warier of relying on scaling relations based on bulge properties, in view of the poor link between BHs and pseudo bulges 
\citep{Kormendy_2013_ARAA_51_511}. Figure~\ref{fig:FR II_BH} shows the Owen-Ledlow relation transformed into the reference frame of BH mass,

\begin{equation}
\log\, \left(\frac{L_\mathrm{1.4}}{{\rm W\,Hz^{-1}}} \right) = 1.35\log\, \left(\frac{M_\mathrm{BH}}{M_\odot} \right) +12.9, 
\end{equation}

\noindent
where we have converted $R$-band absolute magnitude to BH mass using the tight (scatter 0.30~dex) correlation between BH mass and $K$-band luminosity and $V-K=3.0$\,mag \citep{Kormendy_2013_ARAA_51_511}, together with $V-R=0.61$\,mag appropriate for ellipticals \citep{Fukugita_1995_PASP_107_945}.

In the new diagram comparing radio luminosity versus BH mass, the FR~I--FR~II dichotomy is largely restored, although the introduction of the spiral hosts blurs the sharp boundary between the two types of radio sources as traditionally defined by elliptical hosts. It is possible that the FR~I--FR~II boundary depends on the morphological type, such that for a given BH mass spiral hosts generate systematically less radio power than their elliptical host counterparts. This can explain the paucity of FR~I sources currently found among the new population of spiral hosts. The average radio power of FR~Is is $\sim 2$ orders of magnitude weaker than that of FR~IIs. Given the typical flux density of 50\,mJy of the FR~IIs currently detected in disk galaxies, all else being equal the FR~I counterparts would have flux densities of merely 0.5\,mJy, which would place them below the detection limit of FIRST, which is 1\,mJy for individual sources and even worse for extended radio lobes \citep{Becker_1995_ApJ_450_559}. Deeper, higher-resolution observations may yet reveal a more extensive population of FR~I sources hosted in spiral galaxies. Indeed, \cite{McCaffrey_arXiv_2022} detected complex, sub-galactic radio structures in radio-quiet quasars that are possibly FR~I radio lobes, although little is known about the properties of the host galaxies.

\cite{Ghisellini_2001_AAP_379_L1} suggested that the division between the two types of radio sources may be associated with a transition in the accretion mode, from a radiatively inefficient, advection-dominated accretion flow \citep{Narayan1994} in FR~Is to the standard geometrically thin, optically thick accretion disk \citep{Shakura_1973_AA} in FR~IIs. The transition occurs at a critical ratio $\alpha\equiv P_\mathrm{jet} /L_\mathrm{Edd} \approx 10^{-3}-10^{-2}$, where the Eddington luminosity of the BH $L_\mathrm{Edd}\equiv 1.3 \times 10^{38} \,(M_\mathrm{BH}/M_\odot)\, \mathrm{erg \, s^{-1}}$ and jet power \citep{Cavagnolo_2010_ApJ_720_1066}

\begin{equation}
\log\, \left(\frac{P_\mathrm{jet}}{P_1} \right) = (0.75\pm 0.14) \log\, \left(\frac{P_\mathrm{radio}}{P_2} \right) + (1.91\pm 0.18),
\end{equation}

\noindent
with $P_1 = 10^{42}\, \mathrm{erg \, s^{-1}}$, $P_2 = 10^{40} \, \mathrm{erg \, s^{-1}}$, $P_\mathrm{radio}\equiv \nu L_\mathrm{\nu}$, and $\nu=1.4\,\mathrm{GHz}$. Figure~\ref{fig:FR II_BH} shows that the traditional FR~I--FR~II dichotomy of elliptical hosts corresponds to $\alpha \approx 0.005$, consistent with the notion that the FR dichotomy arises from the transition of accretion modes. If a similar dichotomy exists in disk galaxies, we speculate that it would occur at an even lower value of $\alpha$. We note that the FR dichotomy in disk radio galaxies appears less distinct than that for the elliptical galaxies from \cite{Owen_1994}. We suspect that this is partly because our sample, owing to detection limits, lacks FR I objects with low radio luminosities. It is also possible that the dichotomy is intrinsically not sharp and breaks down at low radio power \citep{Best_2012_MNRAS, Whittam_2022_MNRAS} since the original \cite{Owen_1994} sample is affected by Malmquist bias.

\begin{figure*}[]
\plotone{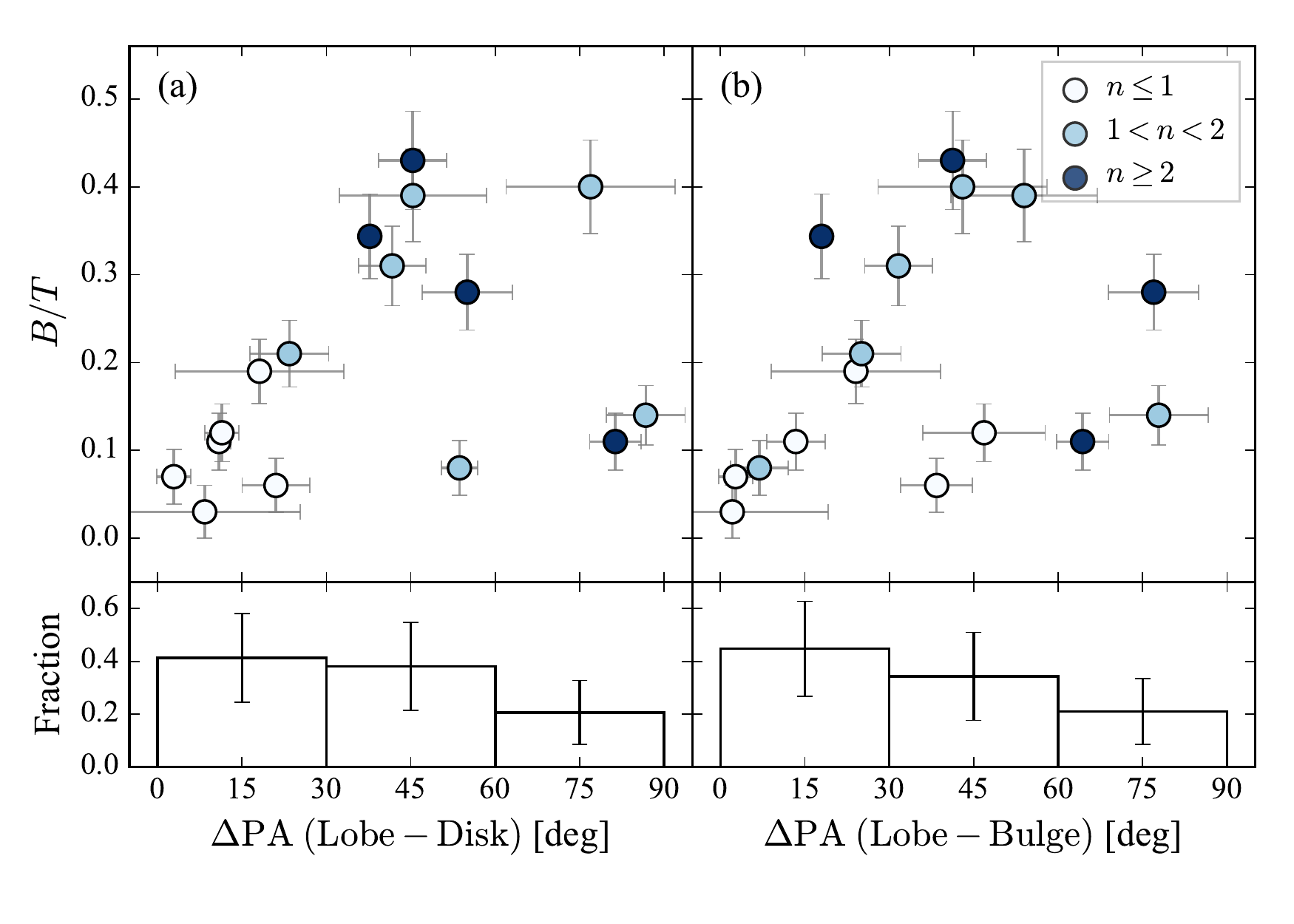}
\caption{The relative position angle ($\Delta$PA) between the radio lobes and the minor axes of the host galaxy (a) disk and (b) bulge. The upper panels show the correlation between $\Delta$PA and bulge-to-total ratio ($B/T$), where the color code indicates bulge \sersic \ index ($n$). The lower panels give the distribution of $\Delta$PA in bins of $30\degr$.}
\label{fig:PA}
\end{figure*}

\subsection{Misalignment Between Radio and Optical Axes}
It has long been realized that the orientation of radio lobes is not preferentially aligned with the minor axis of elliptical galaxies \citep{Birkinshaw_1985_ApJ}. \cite{Browne&Battye_2010_ASPC} found that the alignment is related to the ratio of radio to optical flux, with a higher probability of finding alignment in objects with lower ratios. Adopting their classification criterion, all our sources fall in the high-ratio class. The relative position angles ($\Delta$PA) between radio lobes and the minor axis of galactic disks and bulges span a broad distribution, with lobes ranging from aligned (0\degr) to nearly perpendicular (90\degr) relative to the galaxy (Figure~\ref{fig:PA}). Notwithstanding a slight, apparent trend toward alignment directions, after accounting for measurement uncertainties and Poisson noise the distribution of $\Delta$PA is still consistent with being uniform, with $\chi^2=1.2$ for the disk (Figure~\ref{fig:PA}a) and $\chi^2=1.4$ for the bulge (Figure~\ref{fig:PA}b). Since the orientation of the jet is coupled with the spin of the central BH \citep{Blandford_1977_MNRAS_179_433, Mirabel_1999_ARAA_37_409}, our results suggest that the spin of the BH is not necessarily aligned with the large-scale angular momentum of the host galaxy. 

Although $\Delta$PA is randomly distributed for the whole sample, it appears to be mildly correlated with bulge prominence or bulge type. Host galaxies with the least conspicuous ($B/T<0.2$), low-\sersic\ index ($n\leq 1$) bulges have radio jets that tend to show alignment with the disk and bulge minor axis. In other words, the spin of the central BH of galaxies that evolved secularly through internal gas inflows instead of via mergers may bear the imprint of the angular momentum of the large-scale disk.

\bigskip
\bigskip
\bigskip
\bigskip
\section{Discussion}\label{sec:correlation}
\label{sec:discuss}

Although historically double-lobed radio sources have been found exclusively in giant elliptical galaxies \citep{Best_2005_MNRAS_362_25}, to date no clear consensus exists as to how such an association arises. \cite{Wilson_1995_ApJ_438_62} posited that the galaxy-galaxy mergers that classically are invoked to form ellipticals (e.g., \citealt{Toomre1972}) may also be conducive to spinning up the central BH, a prerequisite for producing strong radio jets if they are powered by BH spin \citep{Blandford_1977_MNRAS_179_433}. This popular explanation has lost some persuasion, however, because rapidly spinning BHs evidently can inhabit galaxies of diverse morphological types, not only ellipticals \citep{Reynolds_2021_ARAA_59}.

This study adds a new twist to the narrative. Analyzing HST images acquired as part of the Zoo Gems project \citep{Keel_2022_AJ_163_150}, we demonstrate unambiguously that at least some double-lobed radio sources can originate from late-type galaxies. Unlike previous reports of a similar nature that were largely based on heterogeneous ground-based images (as summarized in Section~1), the evidence introduced here is incontrovertible because of the benefits afforded by high-resolution HST images of uniform quality. Besides having been systematically selected \citep{Banfield_2015_MNRAS_453_2326}, our final high-confidence sample is twice as large as all previously studied cases of a similar nature combined (Table~5). Our quantitative analysis firmly establishes the late-type morphology of the host galaxies through multiple lines of evidence, including the direct detection of spiral arms, large-scale dust lanes, which are likely edge-on projections of spiral arms, high ellipticity, low global \sersic\ index, low concentration, and, of course, low bulge-to-total ratio. The bulges are not merely modest, but, judging from their low \sersic\ indices and low effective surface brightnesses relative to the Kormendy relation, they can be categorized as pseudo bulges, not classical bulges. These characteristics are extraordinary, but not unprecedented, for the hosts of extended radio sources. The closest analog is J2345$-$0449, a galaxy with a spectacular 1.6 Mpc-scale ``double-double'' radio jet that hosts a pseudo bulge ($n_{\rm bulge} \approx 1$) with $B/T \approx 0.14-0.18$ \citep{Bagchi_2014_ApJ_788_174}. \cite{Singh_2015_MNRAS_454_1556} suppose that radio galaxies with late-type optical morphology may originate from ellipticals that recently acquired a disk component after merging with a disk galaxy. We do not believe that this explanation is viable. An elliptical galaxy that swallows a gas-rich companion---the nearby example of Centaurus~A comes to mind \citep{Baade1954, Graham1979}---can gain a disk, but it cannot lose a gigantic bulge. The sources discussed in this study, which, unlike the majority of previous examples in the literature, have highly robust morphologies and structural parameters derived from high-quality HST images, decidedly do {\it not} have a substantial classical bulge component.

The unusual nature of the above-mentioned characteristics becomes even more acute when we consider the fact that the majority of the spiral hosts have very large stellar masses. More than 60\% (11/18) of our main HST sample has $M_* \gtrsim 10^{11}\,M_\odot$, and within the literature sample, the percentage is similar (56\% or 5/9; Table~5). Curiously, despite the clear presence of spiral structure in their disks\footnote{This is true at least for the main Zoo Gems sample, which enjoys the benefits of HST imaging that the literature sample, apart from $0313-192$ \citep{Keel2006}, does not.}, the optical colors are consistent with those of passive galaxies on the red sequence \citep{Blanton2009}. The subset of galaxies with $M_* \gtrsim 10^{11}\,M_\odot$ has a median optical color, corrected for redshift and Galactic extinction, of $g-r = 0.84\,$mag for the HST sample and $g-r = 0.73\,$mag for the literature sample. This population of supermassive spiral galaxies is qualitatively reminiscent of the superluminous spirals and lenticulars highlighted by \cite{Ogle_2016_ApJ_817_109, Ogle_2019_ApJS_243_14}. Extremely massive disk galaxies in the nearby Universe may have arisen from gas-rich minor mergers \citep{Jackson2022}, or perhaps from having experienced an anomalously quiet merger history \citep{Jackson2020, Zeng2021}.

If galaxy morphology no longer can be regarded as a unique signpost of a galaxy's ability to generate large-scale radio jets, then what physical parameter is responsible? The key factor seems to be the galaxy's stellar mass, or, equivalently, BH mass, to the extent that the two are closely related (Equation~9). Notwithstanding a few outliers (see below), the vast majority of the radio galaxies have stellar masses $\gtrsim 10^{11}\,M_\odot$ or BH masses $\gtrsim 10^{8}\,M_\odot$. This is a longstanding, familiar result in the context of the standard paradigm that radio galaxies are exclusively ellipticals \citep{Dunlop_2003_MNRAS_340_1095, McLure_2004_MNRAS_351_347, Best_2005_MNRAS_362_25}. The results of this study show that this traditional view must be modified. The hosts of double-lobed radio sources encompass not only elliptical galaxies but also a rare population of spiral/disk galaxies, whose common characteristic is that they, too, have unusually large stellar masses (Figure~\ref{fig:FR II_BH}). Of course, not all massive galaxies produce extended radio lobes, but the probability that they do increases with galaxy stellar luminosity or mass \citep{Scarpa_2001_ApJ_556_749,Best_2005_MNRAS_362_25}. As a class, radio-loud AGNs are predominantly hosted by massive galaxies \citep{Mauch_2007_MNRAS_375_931}. Moreover, among the massive galaxies that launch large-scale radio lobes, their radio luminosity or jet power scales roughly with galaxy stellar mass (Figure~\ref{fig:FR II_BH}), which was already implicit in the Owen-Ledlow relation (Figure~\ref{fig:FR II}). Combining Equations~9 and 10, we note that the jet power is almost exactly linearly proportional to BH mass: $P_{\rm jet} \propto M_{\rm BH}^1$.  Again, we now know that these trends hold irrespective of galaxy morphology. So, too, can be said of the FR~I--FR~II dichotomy, except that in disk galaxies the transition between the two FR types may occur at a lower value of the critical ratio ($\alpha$) between jet power and Eddington luminosity. This remains to be verified with future, deeper observations capable of detecting the expected weaker emission of FR~I sources.

If stellar mass---and, by extension, BH mass---is the primary factor that determines the probability that a galaxy launches radio jets, while, at the same time, it also controls the power of the jet that ultimately gets launched, then perhaps we can understand why extended jet structures are so rarely found in spiral galaxies. The stellar mass function of disk-dominated galaxies is weighted toward substantially lower masses than that of bulge-dominated galaxies, and even more so still when compared to that of ellipticals (e.g., \citealt{Moffett2016}). Thus, extended radio jets are a priori expected to be both rare and weak in spiral galaxies. Even when present, the jets will be more compact, as jet size correlates with jet power (e.g., \citealt{Ledlow2002}). If the jet does not extend significantly beyond the boundaries of the stellar disk, it is likely to be confused with the native synchrotron emission from star formation. For example, the AGN components of the radio emission in nearby Seyfert galaxies have 1.4\,GHz powers $\lesssim 10^{23}\,\mathrm{W\,Hz^{-1}}$ and linear source sizes $\lesssim 10\,$kpc \citep{Ho_Ulvestad_2001, Ulvestad_Ho_2001}.

We close with some cautionary notes and suggestions for future work. Our study hinges on the key assumption that we have identified the correct optical counterpart of the host galaxy of the radio lobes. While we have devoted concerted efforts to cull a sample for which spurious chance alignment is low (Section~3), of course, such statistical arguments cannot completely guarantee that all the associations are real. Two sources in our sample stand out as glaring outliers. As discussed in Section~5.1, J0806+062 has a stellar mass of only $M_* = 6.5\times 10^9\,M_\odot$, and J1328+571, with $M_* = 7.8\times10^8\,M_\odot$, is more extreme still. Low-mass galaxies can host nuclear central BHs \citep{Filippenko&Ho_2003_ApJ, Greene&Ho_2004_ApJ, Greene_2020_ARAA_58_257}, and a minority even have radio cores and compact jets (e.g., \citealt{Greene2006, Wrobel2006}), but not classical double-lobed radio structures. Although J0806+062 and J1328+571 have the lowest radio powers in the sample ($L_{1.4} = 2\times 10^{23}$ and $4 \times 10^{22}$\,W\,Hz$^{-1}$), their radio sources, while also small compared to the rest ($D = 24$ and 8\,kpc), are clearly not confined to the nucleus (Figure~1). Follow-up observations are urgently needed to verify the reality of these two baffling sources.

The preceding discussion often presupposes that BH mass truly traces galaxy stellar mass. While this has been established for nearby inactive galaxies for which BH mass can be measured by dynamical methods and for active galaxies for which BH mass can be estimated through their broad emission lines (see reviews in \citealt{Kormendy_2013_ARAA_51_511, Greene_2020_ARAA_58_257}), we acknowledge the inherent uncertainty and ambiguity of applying the $M_{\rm BH}-M_*$ relation to our sample. Which $M_{\rm BH}$--host galaxy scaling relation, if any, is appropriate for these unusual galaxies? An important next step should secure more complete and higher quality optical spectra of the candidate optical counterparts, to fully delineate the ionization source of their nuclei. Are they all AGNs? If so, what type? X-ray observations \citep[e.g.,][]{Mirakhor_2021_MNRAS} would be particularly helpful to disentangle possible confusion arising from dust obscuration or contamination by stellar energy sources. The present sample was originally selected from the FIRST survey, whose sensitivity ($\sim 1\,$mJy) and angular resolution ($\sim 5\arcsec$) can be vastly improved with follow-up observations with the Jansky Very Large Array.

\section{Summary}

We report the existence of disk galaxies as hosts of double-lobed radio sources, using high-resolution optical (F475W) HST images acquired as part of the Zoo Gems program, which targeted 32 sources originally selected from the Radio Galaxy Zoo project that identified double-lobed radio sources associated with disk galaxies from cross-matching the FIRST survey and SDSS. To examine the fidelity and reliability of the physical association between the apparent optical counterparts and radio lobes, we systematically calculate the probability of chance alignment. For a subset of 18 high-confidence objects for which chance alignment is unlikely, we derive the optical morphologies and global and bulge structural parameters, which are combined with the physical properties of the host and radio sources to understand the nature of jet production.

Our main results are as follows:

\begin{itemize}
\item The host galaxies have unambiguous disk-dominated morphologies, as judged by the presence of spiral arms, large-scale dust lanes among the edge-on systems, and low global \sersic\ indices and optical concentrations. Two-dimensional image decomposition yields bulge-to-total light ratios of $B/T = 0 - 0.43$, with a median value of 0.13. The bulges have low \sersic\ indices (median $n_{\rm bulge} = 1.4$) and low effective surface brightnesses that are consistent with pseudo bulges.

\item Despite the obvious morphological and structural properties of late-type galaxies, the majority of the hosts have very large stellar masses (median $M_* = 1.3\times 10^{11}\,M_\odot$) and red optical colors (median $g-r = 0.69\,$mag), consistent with massive, quiescent galaxies on the red sequence. A literature sample of nine radio sources previously found to be hosted by disk galaxies shares strikingly similar characteristics.

\item As with the dominant population of elliptical radio galaxies, among disk radio galaxies the jet power scales with stellar mass. In terms of black hole mass, $P_{\rm jet} \propto M_{\rm BH}^1$.

\item Elliptical radio galaxies display a dichotomy in the $P_{\rm jet}-M_{\rm BH}$ plane, such that sources with an FR~II radio morphology produce systematically more powerful jets than FR~I sources of the same black hole mass. The separation line corresponds to $P_\mathrm{jet}/L_\mathrm{Edd} \approx 0.005$, which may be related to a transition in accretion mode from a standard disk to a radiatively inefficient accretion flow. Nearly all of the currently known disk radio galaxies are FR~II sources. We suggest that among the new population of disky hosts the critical threshold for the accretion mode transition occurs at a lower value of $P_\mathrm{jet}/L_\mathrm{Edd}$, which has yet to be reached by the sensitivity of the current radio observations.

\item The axis of the radio jets is uncorrelated with the minor axis of the host galaxy or its bulge, although galaxies with the smallest bulges ($B/T < 0.2$, $n_{\rm bulge} \le 1$) may show a mild preference to be aligned with the jet, suggesting that the angular momentum on nuclear scales bears the imprint of the angular momentum of the large-scale galactic disk, plausibly as a consequence of gas inflows through secular evolution.
\end{itemize}

\begin{acknowledgments}
We are grateful to the anonymous referee for helpful comments. We thank Subo Dong and Kejia Lee for helpful discussions, and Hua Gao for advice on CGS data. We received financial support from the National Science Foundation of China (11721303, 11991052, 12011540375, 12192220, 12192222) and the China Manned Space Project (CMS-CSST-2021-A04, CMS-CSST-2021-A06). This research is based on observations (program 15445) made with the NASA/ESA Hubble Space Telescope obtained from the Space Telescope Science Institute, which is operated by the Association of Universities for Research in Astronomy, Inc., under NASA contract NAS 5–26555. We made use of the NASA/IPAC Infrared Science Archive, which is funded by the National Aeronautics and Space Administration and operated by the California Institute of Technology.

\software{
{\tt Astrolib PySynphot} \citep{2013ascl.soft03023S}, \texttt{Astropy} \citep{Price-Whelan_2018_AJ_156_123, Robitaille_2013_AAP_558_A33}, \texttt{CASA} \citep{McMullin_2007}, \GALFIT \ \citep{Peng_2002_AJ_124_266,Peng_2010_AJ_139_2097}, \texttt{Numpy} \citep{harris2020array}, \texttt{Photutils} \citep{larry_bradley_2020_4044744}, \texttt{reproject} (\url{https://reproject.readthedocs.io/en/stable/index.html}), \texttt{yaImpute} \citep{Crookston_2008}}
\end{acknowledgments}

\appendix

\section{Images of the low-confidence sample}
We present the images of the low-confidence sample in Figure~A1. The radio lobes and optical galaxies have a considerable probability of being chance-aligned, given the large offsets between galaxies and radio centers. However, some may also have genuine associations because our probability estimation of chance alignment is conservative. Moreover, we have not exploited every piece of information to constrain the probabilities. Future observations, including optical spectroscopy, X-ray observations, and radio observations of higher angular resolution and better sensitivity are needed to confirm their associations.

\begin{figure*}[]
\figurenum{A1}
\includegraphics[width=1\textwidth]{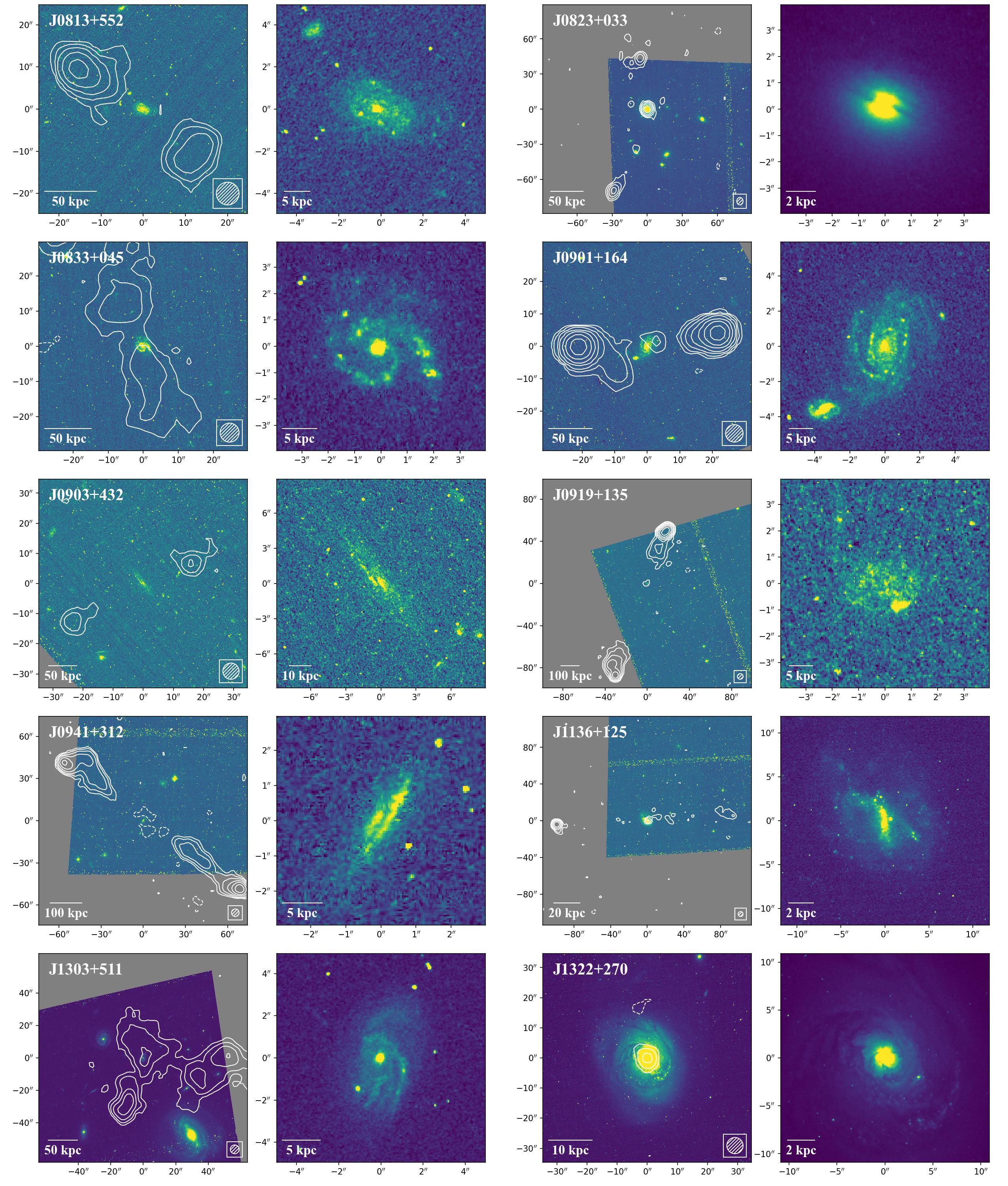}
\caption{HST F475W images of the 14 low-confidence sources in our sample, overlaid with FIRST 1.4\,GHz contours in the left panel and zoomed-in to highlight the optical morphology of the galaxy in the right panel. All images are centered on the galaxy with north up and east to the left. The restoring beam of the radio map is depicted as a hatched ellipse on the lower-right corner. Radio contours are ($-3, 3, 6, 12, 24, 48, 96, ...$) $\times$ rms of each image, where the values of the rms are listed in Table~3. The scale bar in the lower-left corner of each panel indicates the proper distance at the redshift of each object.}
\end{figure*}
\begin{figure*}[h]
\figurenum{A1}
\includegraphics[width=1\textwidth]{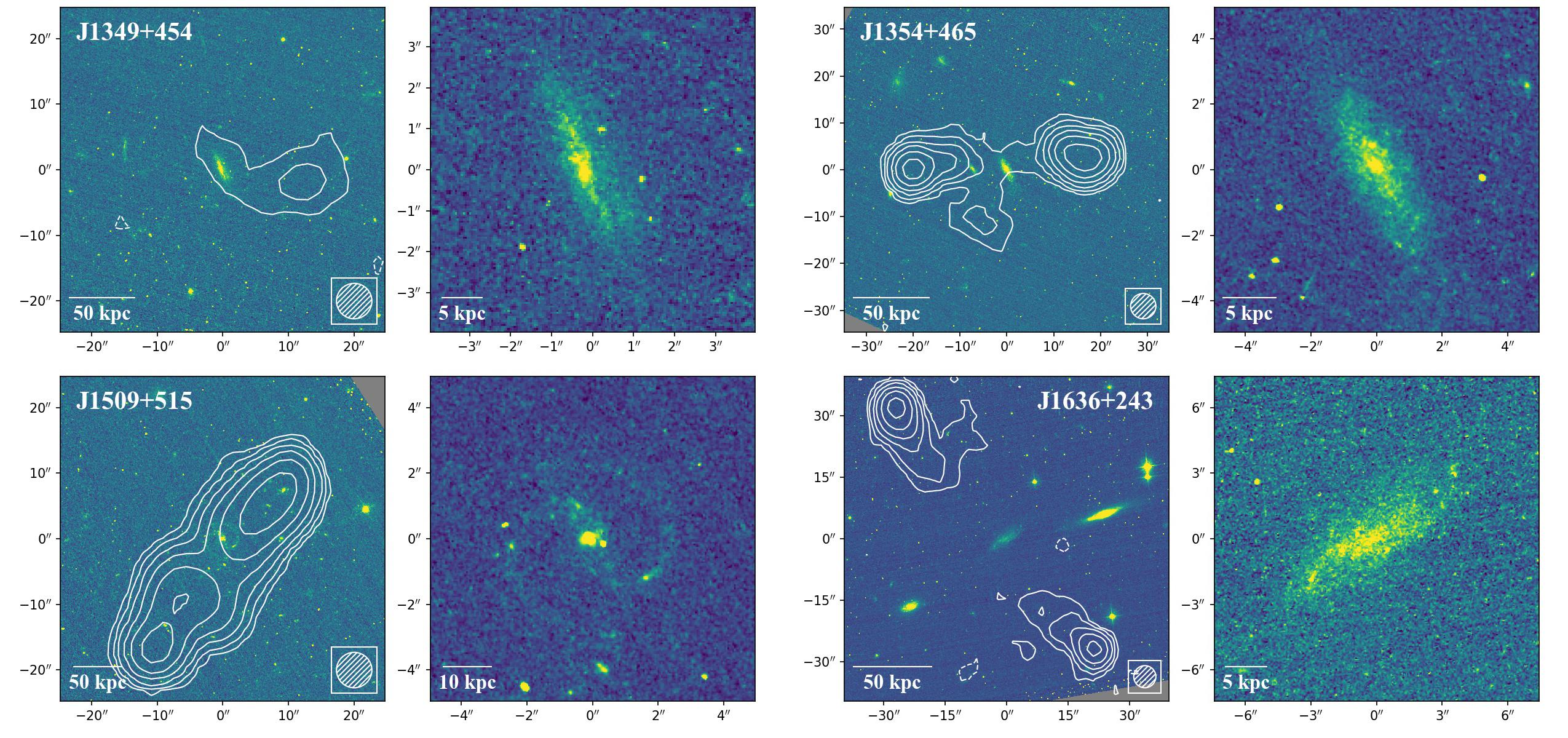}
\caption{\textit{(continued)}}
\end{figure*}

\section{Notes on Individual Sources}
\label{sec:appendix}

\label{JDS44T010}{J0209$+$075}: This is a face-on galaxy with a luminous bulge and loose spiral arms, which harbor several star-forming clumps. The southwestern arm looks disjoint from the galaxy, indicating a low stellar density in the inter-arm region. The bright point-like source northwest of the galaxy is a foreground star identified by SDSS.

\label{JDS43Y010}{J0219$+$015}: This local ($z_\mathrm{spec}=0.04$) galaxy has a fairly prominent bulge and prominent dust structures. The radio source is classified as winged or X-shaped \citep{Yang_2019_ApJS_245_17}.

\label{JDS44I010}{J0802$+$115}: We do not attempt to decompose this interacting system, which has an asymmetric light distribution and long tidal tails. 

\label{JDS44J010}{J0806$+$062}: This galaxy shows several spiral arms and a luminous bar-like region. A galaxy to the northeast with $z_\mathrm{phot}=0.11\pm 0.08$ might be a genuine companion. A star to the east of the galaxy with $m_g=20.7 \,\mathrm{mag}$ has been identified by SDSS and Gaia\footnote{https://irsa.ipac.caltech.edu/cgi-bin/Gator/nph-dd} \citep{Gaia_DR3}. 

\label{JDS45G010}{J0813$+$552}: We assigned the galaxy to the low-confidence sample because it is too faint ($m_g=20.4$\,mag), and the number density at that magnitude is large. In addition, we detect a few sources to the north of the galaxy that might be candidates for the host.

\label{JDS44G010}{J0823$+$033}: This is a complicated case because the double radio lobes have a wide opening angle relative to the galaxy. Our chance-alignment analysis is invalid for this complicated object, and thus we cannot confidently determine the association between the galaxy and the radio lobes. 

\label{JDS44K010}{J0832$+$184}: This galaxy has one of the most prominent bulges in our sample ($B/T=0.4$), and it exhibits a large degree of lopsidedness, suggesting possible signs of interactions. The northern radio lobe is much brighter than the southern one.

\label{JDS45B010}{J0833$+$045}: This object has a high probability of chance alignment because the galaxy is faint and off-center from the radio lobes.

\label{JDS45I010}{J0847$+$124}: This nearly edge-on galaxy has a long dust lane with complicated substructures. The double radio lobes are asymmetric; while the northern lobe is luminous, the southern one is only barely detected. A radio core coincides with the galaxy.

\label{JDS41L010}{J0855$+$420}: This edge-on late-type galaxy, bisected by a prominent, $\sim 10$\,kpc dust lane, has a large stellar mass ($M_*=1.2\times 10^{11}\,M_\odot$). The radio structure has an FR~I morphology because the hotspots are much closer to the galaxy than the lobes. The double lobes have a total length of $\sim 400$\,kpc and a width of $\sim 100$\,kpc (projected). The lobes have sinuous shapes near the hotspots. 

\label{JDS45W010}{J0901$+$164}: A massive spiral with $M_* = 8\times 10^{10}\,M_\odot$, this galaxy has prominent arms and star-forming clumps. The eastern radio lobe has a tail to the southwest instead of toward the direction of the other radio lobe. We crossmatched the peak coordinate of the tail with the AllWISE catalog \citep{Wright_2010_AJ_140_1868} but found no source. A spiral galaxy with $z_\mathrm{phot}=0.22\pm 0.05$ and a prominent bar is located southeast of the main object. Another faint object with $z_\mathrm{phot}=1.35\pm 0.25$ is located to the northwest of the main object and close to the center of the radio contours. To obtain a conservative estimate of the chance-alignment probability, we regard this faint object as a possible radio source.

\label{JDS45L010}{J0903$+$432}: This is a faint ($m_g=20.7$\,mag), very massive ($M_*=2\times 10^{11}\,M_\odot$), edge-on galaxy with a prominent dust lane. The radio lobes are perpendicular to the galaxy plane. It has a high probability of being chance-aligned with the radio lobes because the number density of galaxies of this magnitude is large. 

\label{JDS43Z010}{J0914$+$413}: The galaxy is bisected by a prominent, $\sim 10$\,kpc dust lane. The complicated dust lane introduces significant uncertainties to the bulge-to-disk decomposition. The northern lobe has a second peak with an intensity half of that of the main one, but no counterpart is found in the HST image and AllWISE catalog. In addition to the two lobes, a radio core coincides with the galaxy.

\label{JDS45H010}{J0919$+$135}: This is likely a chance-aligned case because the galaxy is too faint ($m_g=21.5$\,mag) and significantly offset from the center of the radio lobes ($r_\mathrm{offset}=18\farcs1$). Moreover, many other objects are detected near the center of the radio lobes that may be candidate hosts.

\label{JDS47L010}{J0926$+$465}: This is an edge-on galaxy with a dust lane $\sim 20$\,kpc long. In addition to the two lobes, faint radio emission also coincides with the galaxy.

\label{JDS44R010}{J0941$+$312}: We classified this faint edge-on galaxy as a low-confidence object. However, it is located in the center of radio emission, and no other sources are detected nearby. The radio lobes have an angular size of $\sim 2\farcm5$. If they are associated with the galaxy, which has $z_\mathrm{phot}=0.366\pm 0.037$, the physical dimension would be $\sim 1.4$\,Mpc, similar to J2345$-$0449 \citep{Bagchi_2014_ApJ_788_174}.

\label{JDS47H010}{J0956$+$162}: This edge-on galaxy with a prominent ($\sim 15$\,kpc) dust lane has intense radio emission from the nucleus, with peak intensity reaching $\sim 1.1$\,mJy\,beam$^{-1}$. The two radio lobes are roughly equidistant from the center of the galaxy. We find an object with $z_\mathrm{phot}=0.46\pm 0.08$ in the direction of the eastern lobe, $\sim 2\arcsec$ offset from the radio peak; it is unlikely to be associated with the radio lobe because of its small size ($0\farcs91 \pm 0\farcs04$) and faintness ($m_{\rm F475W}=22.5 \pm 0.1 \, \mathrm{mag}$).

\label{JDS44P010}{J0958$+$561}: The galaxy has a prominent bulge and an extended disk. Its nuclear region is complicated in the HST image. The {\tt GALFIT} residuals reveal a compact source near the center, which might be a candidate secondary nucleus (see Figure~\ref{fig:galfit_all}).  In addition to the two lobes, a radio core coincides with the galaxy.

\label{JDS45T010}{J1128$+$241}: This is a nearly face-on spiral galaxy with obvious star-forming clumps along its arms, rendering its optical color ($g-r = 0.46$\,mag) bluer than those of others in the sample.

\label{JDS47K010}{J1136$+$125}: This is a low-mass galaxy ($M_*=2\times 10^9\,M_\odot$) with a peculiar morphology and distortions in its outer regions. It has a star formation rate of $0.4\,M_\odot\,\mathrm{year}^{-1}$ according to GSWLC-2. We do not consider it a likely host for the radio source. We note that there is a bright point source with $m_\mathrm{F475W}$ = 22.1\,mag located only 1\arcsec\ west of the galactic center, which should be investigated further as a possible quasar candidate responsible for the radio lobes.

\label{JDS45E010}{J1303$+$511}: The radio morphology is too complicated to interpret, and thus we do not attempt to link it to the optical galaxy.

\label{JDS45A010}{J1322$+$270}: This is a nearby galaxy without detected radio lobes in the FIRST images. We find no reports of radio lobes in the literature either.

\label{JDS45V010}{J1328$+$571}: This low-mass ($M* = 8\times 10^8\,M_\odot$), low-redshift ($z_\mathrm{phot}=0.032\pm 0.029$) galaxy has many star-forming regions, at least two prominent arms, and a strong bar. With an estimated BH mass of $\lesssim 10^5\,M_\odot$, it qualifies as one of the very few intermediate-mass BHs known to have strong radio emission, and the first of its kind to have a double-lobed, extended jet structure. \cite{Greene_2020_ARAA_58_257} suggest that intermediate-mass BHs may not have sufficient time to sink into the galactic center by dynamical friction. This may explain why the center of the radio lobes is offset from the galactic nucleus by $\sim 4\farcs0$ ($\sim 2.5$\,kpc).

\label{JDS47J010}{J1349$+$454}: We assign this galaxy to the low-confidence sample because only one lobe is detected. It does not satisfy our sample definition of double-lobed radio sources. Further observation may help to verify the existence of the other lobe.

\label{JDS45Z010}{J1354$+$465}: This is a late-type galaxy with a stellar mass of $5\times 10^9 \,M_\odot$. Although classified into the low-confidence sample, it is still a strong candidate for radio association due to its proximity to the center of two radio lobes. Further high-resolution radio observations and optical spectroscopy to secure a better redshift would help to determine the optical-radio association.

\label{JDS44D010}{J1509$+$515}: This galaxy has two grand-design spiral arms, a long bar that reaches $\sim$10\,kpc, and an enormous stellar mass of $5\times 10^{11}\,M_\odot$. We consider it a low-confidence object because it is offset from the center of the radio lobes. However, we regard this galaxy as still a strong candidate, given its unusually large stellar mass.

\label{JDS45J010}{J1633$+$084}:  This massive ($M_* = 2\times 10^{11}\, M_{\sun}$), edge-on galaxy has a prominent, large-scale dust lane. It is quite unexpected that a galaxy this massive would have such a small bulge ($B/T=0.11$), which is likely a pseudo bulge given its very small \sersic\ index ($n_\mathrm{bulge}=0.37$). In addition to the two lobes, a prominent radio core is coincident with the optical galaxy. Since both radio lobes are barely resolved, one might suspect that they are two unrelated point sources instead of radio lobes. However, the chance is extremely small for three unrelated sources to be so well aligned. We estimate the probability that three random radio sources with a projected distance smaller than 50\arcsec \ to be aligned by chance into a straight line with an accuracy of 5\arcsec. Since the FIRST survey has $\sim 220,000$ resolved sources spanning $ 10,000 \, \mathrm{deg}^2$ of the sky with S/N $> 10$ \citep{Banfield_2015_MNRAS_453_2326}, the total number of such alignments is given by \citep{Edmunds_1981_NAT_290_481}: 

\begin{equation}    
N=\frac{2 \pi}{3} \Omega\, d^{3} \,n^{3} \,P,
\end{equation}

\noindent
where $\Omega$ is the sky coverage of the survey, $d$ is the maximum distance between the three sources, $n$ is the source surface density, and $P$ is the alignment accuracy. For this object, we find $N=0.8$. Moreover, the actual probability is likely much lower since we find no optical counterparts for the two radio lobes. Therefore, it is unlikely that the three sources have no physical association, and we consider the two compact, symmetrically aligned radio sources to be radio lobes. 

\label{JDS45F010}{J1636$+$243}: This faint galaxy ($m_g=20.1$\,mag) has a stellar mass of $ 3\times 10^{9} \,M_\odot$. We classify it as a low-confidence source mainly because the number density of galaxies of that magnitude is large.

\label{JDS44Z010}{J1646$+$383}: The galaxy has a prominent dust lane with a peculiar, arc-like shape. Curiously, we find a star projected against a radio contour, located northeast of the galaxy at ($\alpha, \delta) = (16^{\mathrm{h}}46^{\mathrm{m}}29.7^{\mathrm{s}}$, $+38^\circ31{}^\prime36\farcs10$). The parallax measurement from Gaia of $0.44 \pm 0.05$ mas rules out the possibility that it is a quasar, and its SDSS colors ($u-g=1.7$, $g-r=0.6$) are consistent with those of a  K-type star \citep{Covey_2007_AJ}. Thus, it is unlikely to be associated with the large-scale, intense radio emissions. Moreover, its Gaia proper motion of $v_{\rm N,E}=(-3.3 \pm 0.1,2.1 \pm 0.1)$\,mas\,year$^{-1}$ suggests that it is moving from northeast to southwest toward, instead of away from, the center of the radio lobes.

\label{JDS44C010}{J1656$+$640}: This extremely massive ($M_* = 3\times 10^{11}\,M_\odot$) spiral galaxy, classified as star-forming from its narrow emission lines \citep{Baldwin_1981_PASP_93_5}, has a star formation rate of $10 \, M_{\odot}\, \mathrm{year^{-1}}$ according to the GSWLA-2 catalog. A radio core coincides with the galactic nucleus, with an intensity slightly above $3\,\sigma$. We note three other galaxies near the radio center, which are relatively low-redshift objects with $z_\mathrm{phot}<0.3$, not likely distant giant elliptical galaxies.

\label{JDS43V010}{J1721$+$262}: The galaxy has a luminous bulge and an extended disk with spiral arms. A star, identified by SDSS and Gaia, is projected to the south. The two radio hotspots have similar projected distances to the galaxy, with an extended tail emanating from the eastern one. A radio core coincides with the galaxy.

\label{JDS45V010}{J2141$+$082}: The galaxy has an interesting ``eye-like'' shape with a large stellar mass in excess of $ 10^{11}\,M_\odot$.

\section{Two-dimensional Bulge-disk Decomposition Images}
\label{sec:galfit_img}

We present the {\tt GALFIT} bulge-to-disk decomposition results for 16 of the 18 sources in the high-confidence sample in Figure~C1. Two galaxies are omitted: J0802+115 is a highly disturbed interacting system, for which no meaningful decomposition can be done, and J1328+571, which is a dwarf Magellanic spiral that has no detectable bulge.

\begin{figure*}[]
\figurenum{C1}
\centering
\includegraphics[width=0.82\textwidth]{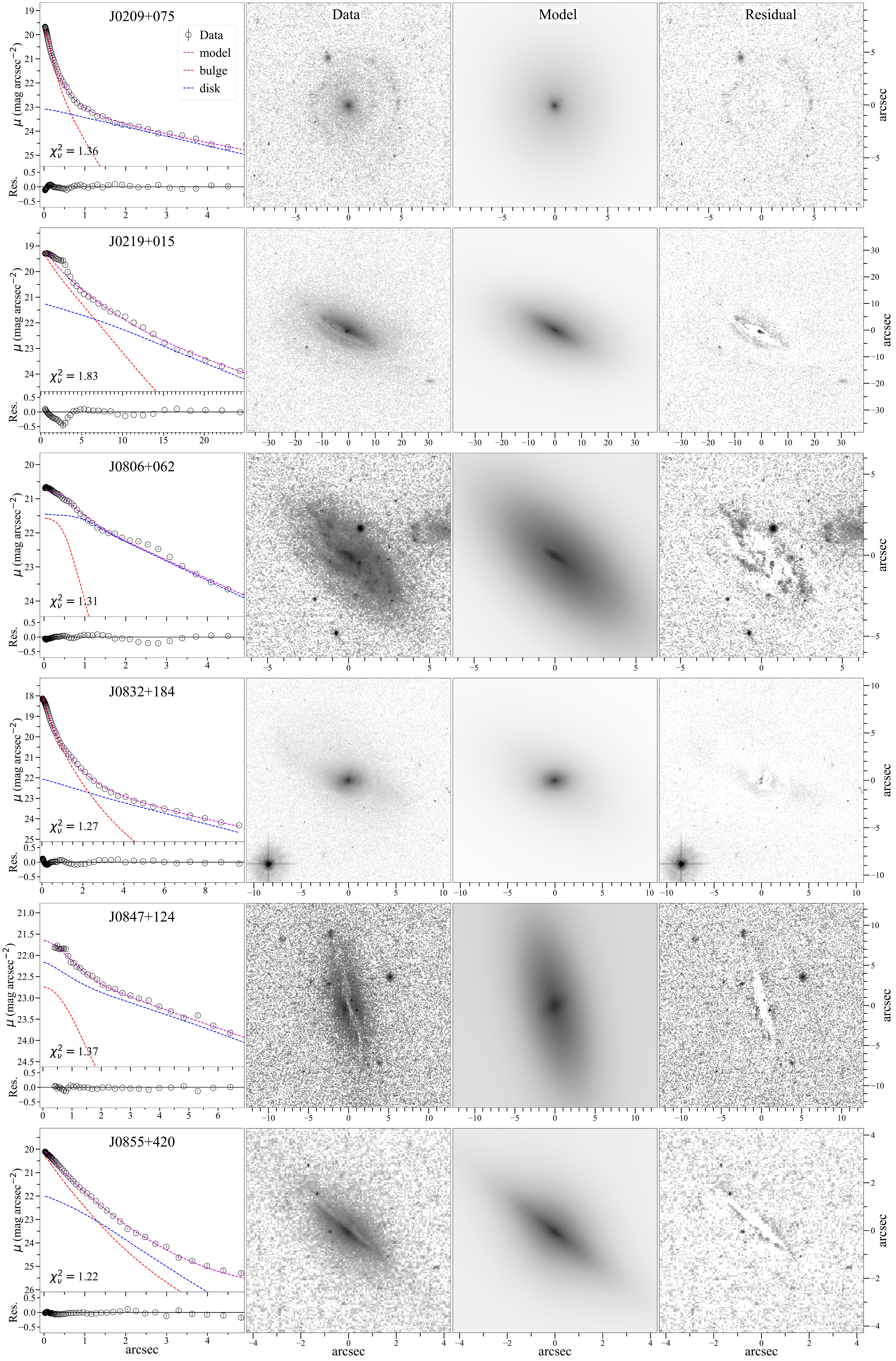}
\caption{Two-dimensional bulge-disk decomposition using \GALFIT\ of 16 of the 18 high-confidence sources in our sample. The left column shows the surface brightness profile of the data (open circles with error bars), \sersic \ bulge component (red), exponential disk component (blue), and total (bulge + disk) model (purple). The $\chi^2_\nu$ from \GALFIT \  is shown in the lower-left corner. The lower panel gives the residuals between the data and the model. The right three columns show the original data, best-fit total model, and residuals, respectively. All images are displayed on a log stretch.
\label{fig:galfit_all}}
\end{figure*}

\begin{figure*}[h]
\figurenum{C1}
\centering
\includegraphics[width=0.82\textwidth]{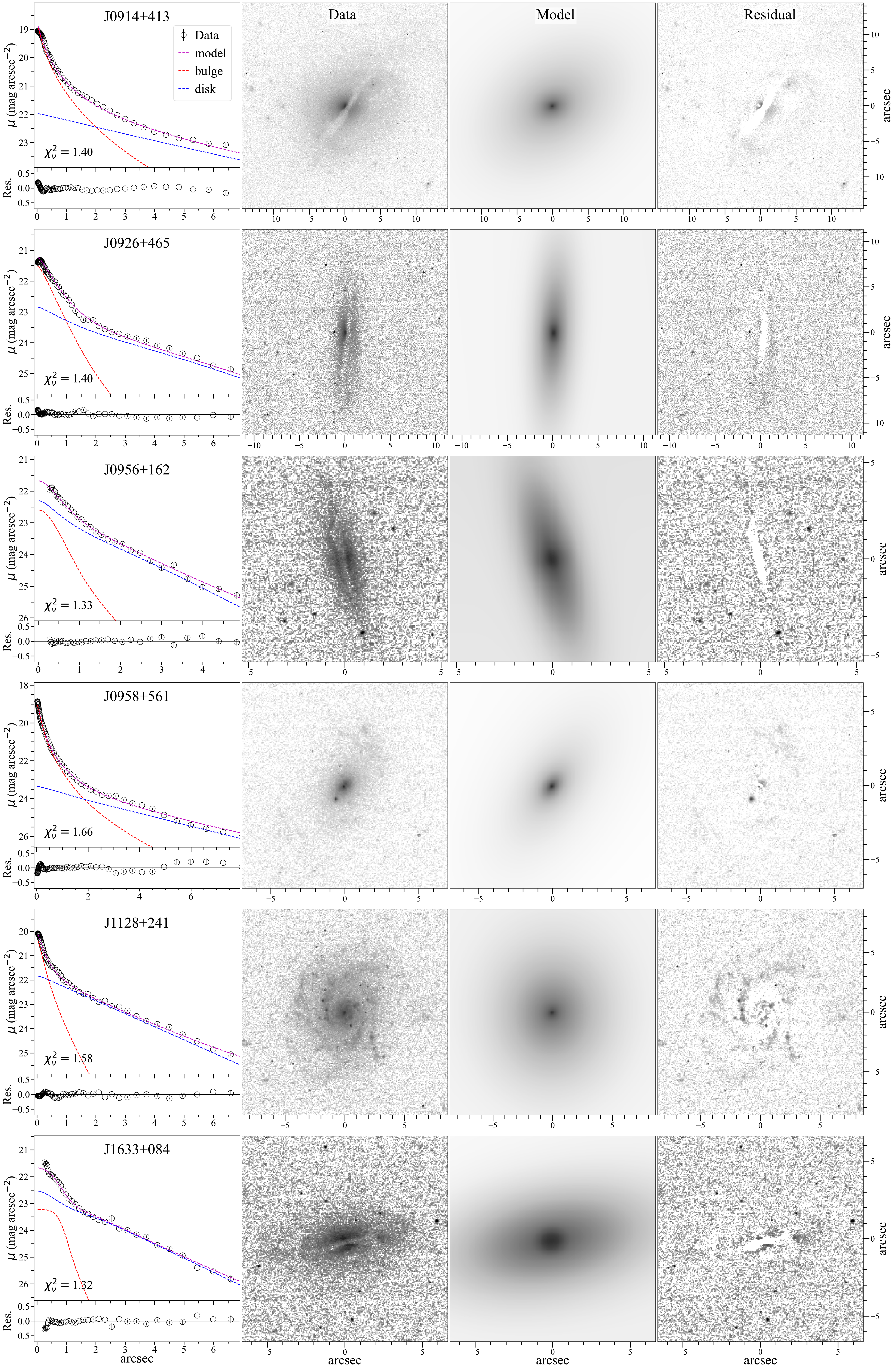}
\caption{\textit{(continued)}}
\end{figure*}

\begin{figure*}[h]
\figurenum{C1}
\centering
\includegraphics[width=0.82\textwidth]{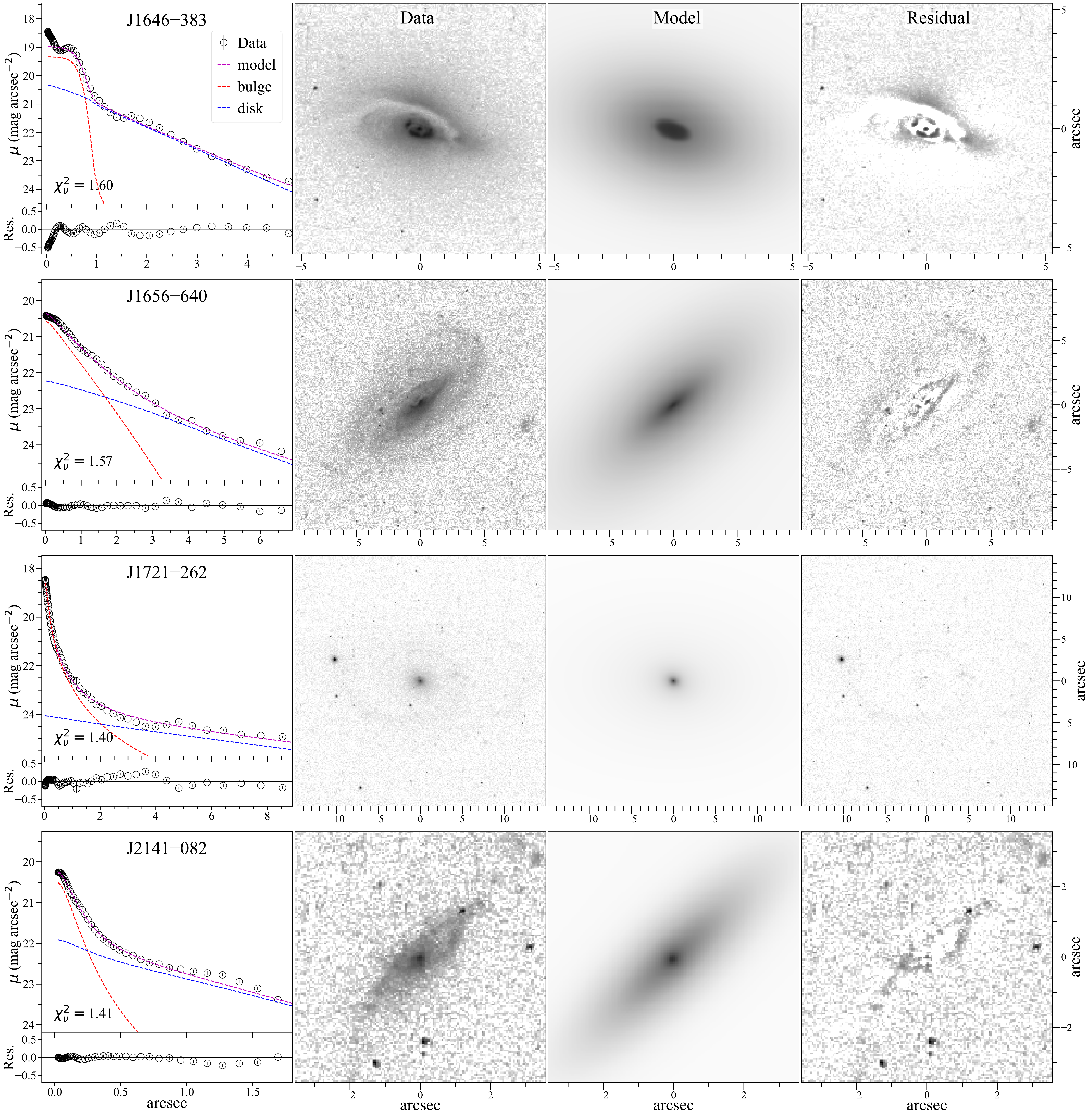}
\caption{\textit{(continued)}}
\end{figure*}

\end{document}